%% file: paper.tex
\documentclass[prd,floatfix]{revtex4}

\usepackage{bm}
\usepackage{epsfig}
\usepackage{amssymb}
\usepackage{amsmath}
\usepackage{slashbox}
\usepackage{calrsfs}
\usepackage{multirow}

\begin{document}

\newcommand{\be}{\begin{eqnarray}}
\newcommand{\ee}{\end{eqnarray}}
\newcommand{\nee}{\nonumber\end{eqnarray}}
\newcommand{\nn}{\nonumber\\}

\title{The non-singlet kaon fragmentation function from $e^+e^-$ kaon production}
\author{S.\ Albino}
\affiliation{{II.} Institut f\"ur Theoretische Physik, Universit\"at Hamburg,\\
             Luruper Chaussee 149, 22761 Hamburg, Germany}
\author{E.\ Christova}
\affiliation{Institute for Nuclear Research and Nuclear Energy of BAS, Sofia 1784, Bulgaria}

\begin{abstract}
We perform fits to the available charged and neutral kaon production data in
$e^++e^-\to K+X$, $K=K^\pm and K^0_S$,
and determine the non-singlet combination of kaon fragmentation functions
$D_u^{K^\pm}-D_d^{K^\pm}$ in a model independent way and without any correlations to the other
fragmentation functions. Only  nuclear isospin invariance is assumed.
Working with non-singlets allows us to include the data at very low momentum fractions,
which have so far been excluded in global fits, and to perform a first NNLO fit to fragmentation functions.
We find that the kaon non-singlet fragmentation function at large $z$
is larger than that obtained by the other collaborations from global fit analysis and differs significantly at low $z$.
\end{abstract}


\maketitle

\section{Introduction}
\label{Intro}

Now that the new generation of high energy experiments with the detection of a final state hadron are taking place,
further tests of QCD and the Standard Model require an accurate knowledge not only of
the parton distribution functions (PDFs) and $\alpha_s(M_Z)$, but equally of the fragmentation
functions (FFs) $D_i^h(z,\mu_f^2)$. These quantities describe the transition of a parton $i$ at factorization scale $\mu_f$
into a hadron $h$ carrying away a fraction $z$ of the parton's momentum or energy in the center-of-mass (c.m.) frame.
Like $\alpha_s(M_Z)$, PDFs and FFs are important quantities because they are universal: according
to the factorization theorem, once they are known at some suitably defined scale $\mu_f=\mu_{f0}$,
they can be calculated at any other scale $\mu_f$ and used in any type of process.
The most reliable way to determine them at a given scale is by fitting to
inclusive single hadron production data in which the fraction $x$ of available momentum or energy in the c.m.\ frame
carried away by the hadron is measured.

While there has recently been quite an extensive study on the PDFs, only
recently have the FFs received more detailed studies,
and it has been recognized that alot of uncertainties appear in their determination.

The most direct way to determine the
FFs is the one-particle inclusive $e^+e^-$ annihilation process:
\be
e^+e^-\to h+X,\qquad h=\pi^\pm,\,K^\pm,\,p/\bar p ...
\ee
Here and from now on we use the shorthand $h^\pm$ to mean either a $h^+$ or $h^-$
is observed (but not both) in a given event.
However, these processes,
being proportional to the square of the
effective electroweak coupling $\hat e_q^2$
of the quark $q$,
determine only the combinations
$D_q^{h^+}+D_q^{h^-}=D_q^{h^+}+D_{\bar q}^{h^+}$,
i.e.\ they
cannot distinguish the quark and anti-quark FFs. In addition, in the limit of massless quarks, they cannot distinguish between
the down-type quark FFs $D_d^h$ and $D_s^h$, which have the same electroweak couplings.
Different assumptions are imposed in order to gain more information about the FFs.
In order to achieve separate determination of
$D_{q}^h$ and $D_{\bar q}^h$, the semi-inclusive DIS $l+N\to l+h+X$ and the one-hadron inclusive production processes $pp\to h+X$ and
$p\bar p\to h+X$ play an essential role. However, in these processes the nucleon structure
is involved, which introduces further uncertainties.

At present several sets of FFs are available in the literature \cite{Albino:2008gy}, such as Kretzer~\cite{Kretzer},
Kniehl-Kramer-Potter (KKP)~\cite{KKP}, Hirai-Kumano-Nagai-Sudoh (HKNS)~\cite{Hirai:2007cx},
de Florian-Sassot-Stratmann (DSS) ~\cite{de Florian:2007hc},
Albino-Kniehl-Kramer (AKK,AKK08) ~\cite{Albino:2005me,Albino:2008fy}, etc.
Two points should be noted about them: 1) in the DSS and HKNS analyses, different relations,
based on theoretical prejudice, between different initial FFs have been imposed,
and 2) there is significant disagreement between the various parametrizations for some of the FFs.
It is not clear how much of this disagreement can be attributed to
the choices of experimental data used by these collaborations and how
much to the choice of the assumptions imposed on the initial FFs.
In this paper we shall consider the possibility of obtaining information about the
FFs directly from experiment, without any assumptions.

Recently, in \cite{Christova:2008te}, we suggested a model independent approach to FFs. We showed
that using only C-invariance of strong interactions,
the difference cross sections between particle and anti-particle production are expressed solely in terms of
non-singlet (NS) combinations of the FFs to any order in perturbative QCD.

There are a number of benefits when performing fits of NS quantities:

i) There are no statistical correlations with gluon FFs, which introduce the largest uncertainties.

ii) In their $\mu_f^2$ evolution they do not mix with other FFs,
so the difference cross sections are independent of the other FFs at all scales.

iii) The NS components do not contain unresummed soft gluon
logarithms (SGLs) at small $z$-values. This allows the use of
measurements at much lower values of $x$ than in global fit
analyses \cite{Albino:2005me,Hirai:2007cx,de Florian:2007hc,Albino:2008fy}, which
would (hopefully) better constrain the NS.
This would provide stronger tests relative to global fits
on the validity of the leading twist calculations at small $x$, where the effects of
higher twists, as well as of quark and hadron masses,
should be most pronounced.

iv) A next-next-to leading order (NNLO) fit of the non-singlet components is
possible, because the perturbative components in the NS sector, namely the splitting and coefficient functions, are known to NNLO.
This is in contrast to global fit analyses where only next-to leading order (NLO) calculations of cross sections are possible at
present.

Note that here and further on we use the notation $x$ for the {\it
measured} fraction of the {\it energy $E$ of the process} carried
away by the observed hadron $h$,  while we use $z$ for the fraction
of the (unobservable) energy $E_p$ of the {\it fragmenting parton}
carried by the observed hadron:
 \be
 x=\frac{2(P^hq)}{q^2}\simeq E^h/E,\qquad z=E^h/E_p,
 \ee
  i.e. $x$ is the
measured quantity, $z$ is the {\it theoretically} QCD-defined
quantity, $E^h$ is the c.m. energy of the observed hadron. In leading order (LO), neglecting transverse momenta and
hadron mass corrections, $x$ and $z$ coincide.

In \cite{Christova:2008te} a model independent approach for
determining NS combinations of FFs was developed. It was shown that
if both charged and neutral kaons are measured in $l+N\to l+K +X$,
in $pp\to K +X$ or in $e^+e^-\to K+X,\, K=K^\pm,K^0$,
SU(2) isospin invariance of strong interactions implies that the cross section differences
$\sigma^{\cal K}$ between the charged and neutral kaons:
\be
d\sigma^{\cal K}\equiv\sigma^{K^\pm}-2\sigma^{K^0_S}\label{calK}
\ee
always determines, without any assumptions about PDFs and FFs, the non-singlet $D_u^{K^\pm}-D_d^{K^\pm}$.

In this paper we apply the model independent approach of
\cite{Christova:2008te} to the available data on $K^\pm$ and $K^0_S$
production in $e^+e^-$ annihilation and determine the kaon non
singlet $D_u^{K^\pm}-D_d^{K^\pm}$. This allows us for the first time
i) to extract $D_u^{K^\pm}-D_d^{K^\pm}$ without any assumptions
about the unfavoured FFs, commonly used in global fit
analysis, ii) to extract $D_u^{K^\pm}-D_d^{K^\pm}$ without any
correlations to other FFs, and especially to $D_g^{K^\pm}$, iii) to
determine $D_u^{K^\pm}-D_d^{K^\pm}$ in a larger region
than in global fits by using all
available data, that is typically in the region $x \gtrsim 0.001$, and iii)
to perform a first NNLO extraction of the FFs.
Including the small $x$ data should also improve the precision of
the FFs at large $z$ since, via the convolution in eq.\ (\ref{genslfacttheo}) below,
all $z$ values in the range $x < z < 1$ contribute,
iv) to perform a first phenomenological test
of recent NNLO calculations and v) to test, at lower $x$ values than before,
the incorporation of hadron mass according the procedure of Ref.\
\cite{Albino:2005gd}, which becomes more important as $x$ decreases.

The rest of the paper is organized as follows. In section
\ref{approach} we describe our approach to charged and neutral kaon production. We show how SU(2) invariance allows
to single out the NS combination of the kaon FFs, and our basic
formula for $e^+ e^-$-kaon production is presented. In section
\ref{choices} we describe our method of analysis and justify the
choice of the parametrizations used. The results of our fits and
the comparison with those obtained from global fits
are discussed in section \ref{results}.
The results are summarized in section \ref{summary}.
Appendix \ref{NNLOHS} outlines our approach for calculating the Mellin transform of harmonic polylogarithms,
which is necessary for the NNLO calculations.

\section{Our formalism}
\label{approach}

In this section we describe our approach for extracting the
kaon non singlet and contrast it to that in global fits.

In general, the factorization theorem implies that any inclusive
hadron production cross section can be written as
\be
d\sigma^h(x,E_s^2)=\sum_i \int_x^1 dz
d\sigma^i\left(\frac{x}{z},E_s^2,\mu_f^2\right) D_i^h(z,\mu_f^2)
+O\left(\left(\frac{1}{E_s}\right)^p\right) \label{genslfacttheo}
\ee
where $E_s$ is the energy scale of the process, $d\sigma^i$ is the process dependent partonic level cross
section for the inclusive production of a parton $i$, determined
fully in terms of perturbatively calculable coefficient functions,
electroweak factors, and of the PDFs for any initial state hadrons,
$\mu_f$ is the factorization scale, and $p\geq 1$.
Note that though formally $d\sigma^i$ is independent of the renormalization scale $\mu$ that
appears as the argument of the running coupling $a_s = \alpha_s /(2\pi)$, it depends on it
when calculated in perturbation theory, further we assume $\mu^2=\mu_f^2$ as usually done.
In LO the measurable quantity $x$ and the QCD variable $z$
usually coincide because $d\sigma^i\left(x/z,E_s^2,\mu_f^2\right) \propto
\delta (z - x)$.

Although the $z$ dependence of the fragmentation
functions $D_i^h(z,\mu_f^2)$ is not calculated perturbatively, QCD determines
perturbatively, via the DGLAP evolution equations,
their $\mu_f^2$-dependence:
\be
\frac{d}{d\ln \mu_f^2}D_i^h(z,\mu_f^2)=\sum_j \int_z^1
\frac{dz'}{z'}P_{ij}\left(\frac{z}{z'},a_s(\mu_f^2)\right)
D_j^h(z',\mu_f^2)
\ee
where $P_{ij}(z,a_s)$ are the perturbatively
calculable splitting functions. In addition, the DGLAP equations allow a choice
of $\mu_f = O(E_s)$ which prevents the large logarithms $\log (E_s/\mu_f)$
from spoiling the accuracy of the perturbative calculations of
$d\sigma^i$. Flavour and charge
conjugation symmetry of QCD allow to combine the quark FFs and quark coefficient
functions into singlets and non singlets, whose advantage is that they
do not mix in their evolution. In this paper we shall deal with non-singlets.

In any kaon-production process, if in addition to the charged $K^\pm$-kaons also the neutral $K_S^0$-kaons are measured,
no new FFs above those used for $K^\pm$ are introduced in the cross section. This is a consequence of SU(2)
invariance of the strong interactions, which relates neutral and charged
kaon FFs:
\be
D_{u,d,s,c,b,g}^{K_S^0}=\frac{1}{2}D_{d,u,s,c,b,g}^{K^\pm}.\label{SU2}
\ee
Then for the difference cross section
$d\sigma^{\cal K}$, eq. (\ref{calK}), we obtain the simple expression:
\be
d\sigma^{\cal K}(x,E_s^2)=\int_x^1 dz
(d\sigma^u-d\sigma^d)\left(\frac{x}{z},E_s^2,\mu_f^2\right)
(D_u^{K^\pm}-D_d^{K^\pm})(z,\mu_f^2) \label{QCDcalcofchKmnK}
\ee
i.e.\ in any inclusive hadron production process
$d\sigma^{\cal K}$ always depends only on one NS combination of FFs, namely
$D_u^{K^\pm}-D_d^{K^\pm}$. This result relies only on SU(2)
invariance for the kaons, eq. (\ref{SU2}), and does not involve
any other assumptions about PDFs or FFs. It holds
in any order in QCD. The
explicit expressions for $d\sigma^{\cal K}$ in $e^+e^-$, SIDIS and
$pp$ scattering were given in Ref.\ \cite{Christova:2008te}.

In this paper, we focus on the most precisely measured and most accurately calculated processes
\be
e^+ e^- \rightarrow (\gamma ,Z)\rightarrow K+X,\qquad K=K^\pm,K^0_S,
\label{e+e-}
\ee
for which eq.(\ref{QCDcalcofchKmnK}) reads:
\be
d\sigma^{\cal K}_{e^+e^-}(x,s)=N_c \sigma_0(s)\,\int_x^1 dz (\hat e^2_u -\hat e_d^2)(s)C_q\left(\frac{x}{z},
\frac{s}{\mu_f^2}, a_s(\mu_f^2)\right)(D_u^{K^\pm}-D_d^{K^\pm})(z,\mu_f^2),
\label{basic}
\ee
where $\sqrt{s}$ is the c.m.energy of the process, $x=2E_h/\sqrt s$,
$\sigma_0=4\pi\alpha_{em}^2/s$ is the Born level cross section for the process $e^+ e^- \rightarrow \mu^+ \mu^-$,
$N_c$ is the number of colours,
and $\hat e_q^2(s)$ is the square of the effective electroweak charge of the quark $q$:
\be
\hat{e_q}^2(s) =\hat e_q^2 - 2\hat e_q
\,v_e\,v_q\,\Re e \,h_Z + (v_e^2 + a_e^2) \, \left[(v_q)^2
+(a_q)^2\right]\, \vert h_Z\vert ^2,
\ee
with $ h_Z = [s/(s-m_Z^2+im_Z\Gamma_Z)]/\sin ^2 2\theta_W$,
$\hat e_q$ the charge of the quark $q$ in units of the proton charge, and
\be
v_e&=&-1/2 +2 \sin^2\theta_W,\quad a_e=-1/2, \nn
v_q&=&I_3^q-2\hat e_q\sin^2\theta_W,\quad a_q=I_3^q, \quad I_3^u = 1/2,
\quad I_3^d = -1/2.
\ee
We set $\mu_f^2=ks$, $k=1$, $1/4$ and 4
to estimate the theoretical error, i.e.
we consider three different choices for $\mu_f$: $\mu_f=\sqrt s /2;\sqrt s $ and $2\sqrt s$.
The energy fraction $z$ is given by $z = 2(P^h.q)/q^2=E_h/E_p $,
$C_q$ is the flavour independent perturbatively calculated quark coefficient function:
\be
C_q(z, \mu_f^2/s, a_s(\mu_f^2))= \delta(1-z)+a_s(\mu_f^2) C_q^{(1)}(z, \mu_f^2/s)+O(a_s^2).
\ee

Eq.\ (\ref{basic}) is our basic formula which we shall use in our fit to determine
$(D_u^{K^\pm}-D_d^{K^\pm})$.

In our analysis we shall use all available $K^\pm$ and $K^0_S$ production data presented
by the different collaborations
TASSO~\cite{Althoff:1984iz}---\cite{Braunschweig:1988hv}, MARK II~\cite{Schellman:1984yz},
TPC~\cite{Aihara:1984mk},
HRS~\cite{Derrick:1985wd}, CELLO~\cite{Behrend:1989ae}, TOPAZ~\cite{Itoh:1994kb},
ALEPH~\cite{Buskulic:1994ft}, DELPHI~\cite{Abreu:1998vq}, OPAL~\cite{Abbiendi:2000cv}---\cite{Abbiendi:1999ry}
and SLD~\cite{Abe:2003iy} at different values of $s$.

\begin{figure}[h!]
\begin{center}
\includegraphics[width=9cm]{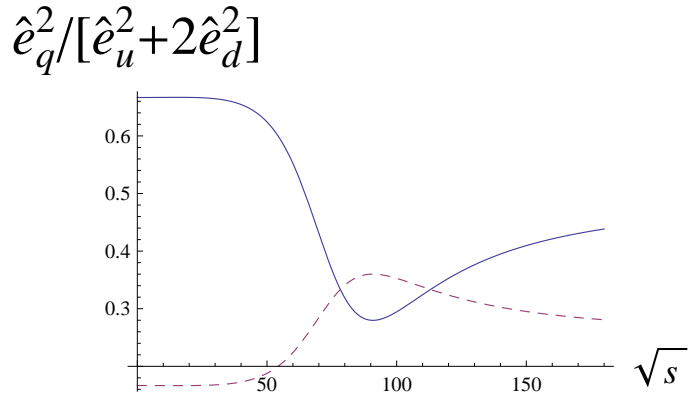}
\caption{The normalized electroweak charges $\hat e_u^2/(\hat e_u^2+
2 \hat e_d^2)$  (full line) and $\hat e_d^2/(\hat e_u^2+ 2 \hat
e_d^2)$  (dashed line) as functions of $\sqrt{s}$.\label{figeued}}
\end{center}
\end{figure}

Experimental data for hadron production (\ref{e+e-})
are commonly presented as normalized to the total hadron cross section
$\sigma_{tot}\simeq \sigma_0\,\sum_q\hat e^2_q$. From eq.(\ref{basic}) it is clear
that the sensitivity of $\sigma^{\cal K}_{e^+e^-}(s)/\sigma_{tot}$ to $(D_u^{K^\pm}-D_d^{K^\pm})$
is determined by the $s$-dependence of $(\hat e^2_u-\hat e_d^2)(s)/\sum_q \hat e_q^2$.
In Fig.\ \ref{figeued} the quantities $\hat e^2_u/\sum_q \hat e_q^2$ and $\hat e^2_d/\sum_q \hat e_q^2$
are shown as functions of $\sqrt{s}$, which demonstrates that
the biggest contribution would come from data away from
the intersections with the $\sqrt s$-axis and the region between them, namely away from
$80 \leq \sqrt{s}\leq 110$ GeV, i.e. most important for our studies would be data for which $\sqrt{s}\lesssim 60$ GeV.
It is unfortunate that at the $Z$-pole $\sqrt s \simeq 91,2$ GeV,
where the most precise and abundant
data exist, the kaon cross section difference normalized to $\sigma_{tot}$
is an extremely small quantity:
$(\hat e^2_u-\hat e_d^2)/\sum_{q=u,d,s}\hat e_q^2= (v_u^2-v_d^2)/[\hat e_u^2+2\hat e_d^2]\simeq -0.081$.

For a large part of the $e^+ e^-$ reaction data for kaon production,
the primary quark (i.e.\ the quark at the electroweak vertex) is ``tagged''.
Experimentally, various techniques are used to achieve this and we refer the reader to the various experimental papers
(but see in particular Refs.\ \cite{Letts:1996sp} and \cite{Abbiendi:1999ry}).
For our calculations, we simply neglect the contributions from all processes except those for
which the primary quark is tagged.
Since this can be achieved by setting the electroweak charges of all quarks to zero except the tagged quark,
the resulting cross section is scheme and scale independent as a physical quantity should be.

We calculate $a_s(\mu^2)=f(L)/(\beta_{0}L)$, where
 $L=\ln \mu^{2}/\Lambda_{\rm QCD}^{2}$ and for $f$ in LO, NLO and NNLO we have:
\be
f_{\rm LO}=1,\qquad f_{\rm NLO}(L)=
1-\frac{\beta_{1}}{\beta_{0}^{2}}\frac{\ln L}{L}
\ee
\be
f_{\rm NNLO}(L)=f_{\rm NLO}(L)+\left(\frac{\beta_{1}}{\beta_{0}^{2}}\right)^{2}
\frac{\ln^2 L-\ln L+
\frac{\beta_{0}\beta_{2}}{\beta_{1}^{2}}-1}{\ln^2 L}.
\ee
The constants $\beta_i$ are given by \cite{Larin:1993tp}
\be
\beta_{0}=\frac{11}{6}C_{A}-\frac{2}{3}T_{R}n_{f}
\ee
\be
\beta_{1}=\frac{17}{6}C_{A}^{2}-C_{F}T_{R}n_{f}-\frac{5}{3}C_{A}T_{R}n_{f}
\ee
\be
\beta_{2}=\frac{2857}{432}C_{A}^{3}+\frac{1}{4}C_{F}^{2}T_{R}n_{f}-\frac{205}{72}C_{F}C_{A}T_{R}n_{f}
-\frac{1415}{216}C_{A}^{2}T_{R}n_{f}
+\frac{11}{18}C_{F}T_{R}^{2}n_{f}^{2}+\frac{79}{108}C_{A}T_{R}^{2}n_{f}^{2},
\ee
where $C_{A}=3$, $C_{F}=\frac{4}{3}$ and $T_{R}=1/2$.
We fix $\Lambda_{\rm QCD}$ = 226 MeV at both NLO and NNLO and for $n_f=5$
This is the value of $\Lambda_{\rm QCD}$ obtained in the CTEQ6.6M PDF extraction \cite{Nadolsky:2008zw}.

\section{Method of analysis of $K^\pm$ and $K^0_S$ data simultaneously}
\label{choices}

Our formalism would be easy if
we had data on $K^0_S$ and $K^\pm$ production at identical values of $x$ and $\sqrt{s}$,
with the cross sections being normalized in the same way. Then
the optimum procedure to constrain the kaon non singlet would be to fit it
to the difference between these data.
However, apart from the $u$ and $d$ quark tagging probabilities from OPAL,
this does not hold for the data in general.
Data on $K^\pm$ and $K^0_S$ are at similar c.m.\ energies $\sqrt{s}$,
but usually at quite different $x$ values.
Therefore we proceed in 4 steps:

1) We combine the measurements on $K^0_S$ into seven energy intervals
$\sqrt{s}=$
12 -- 14.8, 21.5 -- 22, 29 -- 35, 42.6 -- 44, 58, 91.2 and 183 -- 186 GeV and {\it parametrize} the $x$ dependence of the cross section
$d\sigma^{K_S^0}$ for $K^0_S$-production in each interval separately as defined below in eq. (\ref{paramofK0Sdata}).

2) For each interval of $\sqrt{s}$ we {\it calculate}
$d\sigma^{\cal K}$ {\it perturbatively}, using eq.\ (\ref{basic}),
parametrizing the $z$ dependence of the kaon non singlet at a
suitable starting scale $\mu_f=\mu_{f0}$, as described below in eq. (\ref{1stkaonnonsing}).

3) Using (\ref{paramofK0Sdata}) and (\ref{1stkaonnonsing}) we calculate the
charged kaon cross section $\sigma^{K^\pm} $ in each energy interval of $\sqrt{s}$
through the expression:
\be
d\sigma^{K^\pm}_{e^+e^-} = d\sigma^{\cal K}_{e^+e^-} + 2 d\sigma^{K^0_S}_{e^+e^-}.
\ee

4) We fit the parameters in $d\sigma^{K^0_S}$ (as given in eq.\ (\ref{paramofK0Sdata}))
and the parameters in the kaon non singlet (as given in eq.\ (\ref{1stkaonnonsing})) simultaneously
to measurements of charged and neutral kaon production in $e^+ e^-$ reactions.

We believe that the above approach is the optimum one since it involves performing only one fit.

Since the perturbative calculation of the cross section difference $d\sigma^{K^\pm}-2d\sigma^{K_S^0}$
is free of SGLs, it is expected
to be valid at much lower values of $x$ than the perturbative calculations of $d\sigma^{K^\pm}$
and $d\sigma^{K_S^0}$ separately.
Therefore in our fits we vary the lowest value of $x$ that the data can take.
In global fit analyses
the usual minimum bound of $x\geq 0.1$ -- $0.05$ was used, but in general we will include data at lower values.

We parametrize the cross section $d\sigma^{K_S^0}$ as follows:
\be
\frac{d\sigma^{K_S^0}_{e^+e^-}}{dx}(x,s)=\left(N(s)+\frac{\Delta N(s)}{\ln \sqrt{s}}\right) x^{A(s)} (1-x)^{B(s)}
\exp[-c(s) \ln^2 x +d(s) \ln^3 x + e(s)\ln^4 x]
\label{paramofK0Sdata}
\ee
where $N$, $\Delta N$, $A$, $B$, $c$, $d$ and $e$ are seven different parameters
that are fitted to the data in each range of $\sqrt{s}$ separately.
The $\Delta N(s)/\ln \sqrt{s}$ term is motivated by the dependence of the cross section on $\sqrt{s}$ predicted by QCD.
In the case where data of only one $\sqrt{s}$ value exists, namely the data at
$\sqrt{s}=$ 58 and 91.2 GeV, $\Delta N$ is fixed to zero.
Otherwise, note that no QCD input is used for the calculation of the $K_S^0$ production data.
The motivation behind the choice of the parametrization is empirical,
although the $(1-x)^B$ behaviour at large $x$ and the $\exp[-c \ln^2 x]$ at small $x$ also follow from
resummation in perturbative QCD in these respective regions for $\sqrt{s}\gg\Lambda_{\rm QCD}$.

Due to flavour symmetry, $d\sigma^{K^\pm}-2d\sigma^{K_S^0}$ vanishes
whenever the quark at the electroweak vertex is neither a $u$ nor $d$-quark.
Thus, we do not need the $s$, $c$ and $b$-quark tagged data from OPAL that should automatically cancel and
cannot constrain the kaon
non singlet in our approach. However, we shall use the light-quark tagged data, that contain the $u$ and $d$-quarks.
We can parametrize these data directly, but instead
we parametrize $c$ and $b$ quark tagged data as in eq.\ (\ref{paramofK0Sdata}),
and calculate the light quark tagged cross section as
the difference between the untagged cross section and the sum of the $c$ and $b$ quark tagged cross sections.
By including all available heavy quark tagged data in this way, we hope to improve our calculation of the light quark tagged data.
Thus we have 9 parametrized functions in $x$ to describe all the $K_S^0$ data:
seven parametrizations for the untagged data in each $\sqrt s$-energy interval
and 2 for the $c$ and $b$ quark tagged cross sections at $\sqrt{s}=91.2$ GeV.

For the calculation of $d\sigma^{\cal K}$ using eq.\ (\ref{basic}),
we require a parametrization for the kaon non singlet at a starting scale $\mu_f=\mu_{f0}$ which satisfies the
following conditions: It should exhibit the power-like behaviour $z^a$ as $z\rightarrow 0$.
Note that the resummed double logarithmic contribution to the splitting functions
suggests that a Gaussian behaviour in $\ln z$ at small $z$ occurs only for the gluon and singlet FFs \cite{Albino:2004yg} and
we do not assume that this behaviour occurs also for the non singlet.
The FF should also exhibit the behaviour $(1-z)^b$ as z approaches 1.
After trying various parametrization that were in accordance with the above requirements,
we found that the best parametrization, i.e.\ the one that gave a good fit with all parameters
well constrained by the data (meaning that the parameters did not become large), was
\be
(D_u^{K^\pm}-D_d^{K^\pm})(z,\mu_{f0}^2)=n z^a (1-z)^b +n' z^{a'} (1-z)^{b'}.
\label{1stkaonnonsing}
\ee
This parametrization is effectively the same as the one used in the latest global fits in
\cite{de Florian:2007hc,Albino:2008fy}, except that $a' \neq a$ in order to allow a larger function space at small $z$ to be available
to the non singlet.

To be clear,   our main fit (discussed in subsection \ref{mainfit}) which determines the NS $D_{u-d}^{K^\pm}$  proceeds as follows.
We determine  $D_{u-d}^{K^\pm}$ in a simultaneous fit to $K^\pm$ and $K^0_s$ production data --
we fit the $K^0_S$ production data to eq.\ (\ref{paramofK0Sdata}) and we fit the $K^\pm$ production data
to the difference  of eq.\ (\ref{paramofK0Sdata}) (multiplied by 2)
and $d\sigma^{\cal K}$:  $d\sigma^{K^\pm}= 2 d\sigma^{K^0_s}-d\sigma^{\cal K}$,  where $d\sigma^{\cal K}$
is calculated from $D_u^{K^\pm}-D_d^{K^\pm}$ using eq.\ (\ref{QCDcalcofchKmnK}).
Note that if all the $K^0_S$ production data were measured at the same $x$ and $\sqrt{s}$ values, and defined in the same way,
as the $K^\pm$ production data, there would be no need for eq.\ (\ref{paramofK0Sdata}) --- we would simply fit
the theoretical calculation of $d\sigma^{\cal K}$, eq.\ (\ref{QCDcalcofchKmnK}), directly to the measurements of
$d\sigma^{\cal K}$ at each measured $x$ and $\sqrt{s}$ value.
We stress that, despite the theoretical discussion immediately following eq.\ (\ref{paramofK0Sdata}),
the motivation for the parameterization in eq.\ (\ref{paramofK0Sdata}) is  mainly empirical --
as we will see in subsection \ref{KSonly}, such a parameterization describes all $K_S^0$ production data well.
We note, however, that different parameterizations will exist which give an equally good fit
to the $K_S^0$ production data but give slightly different results.
Such a ``parameterization error'' should in any case be less than the errors on the parameters due to the
errors on the measurements.
We also note that a single simultaneous fit of all parameters to all data is the statistically correct approach.
For example, fitting the parameters in eq.\ (\ref{paramofK0Sdata}) to the $K_S^0$ production data
and then, as a separate fit, fitting the parameters in eq.\ (\ref{1stkaonnonsing}) to the $K^\pm$ data only
would not take into account the fact that the fitted values of the the parameters in eq.\ (\ref{paramofK0Sdata})
carry significant experimental errors.

In our perturbative calculations, we choose $\mu_{f0}=\sqrt{2}$ GeV, 5 active flavours $u,d,s,c,b$,
and $\Lambda_{\rm QCD}=226$ MeV.
We perform all calculations in Mellin space since this approach is numerically more efficient
than explicitly performing $x$ space convolutions such as that in eq.\ (\ref{1stkaonnonsing}).

The NNLO perturbative components for the cross section difference can be obtained
using the results of \cite{Rijken:1996vr} for the non singlet coefficient functions
and the results of \cite{Mitov:2006ic} for the difference between the spacelike and timelike non singlet splitting functions.
The former, as well as the spacelike non singlet splitting functions of \cite{Moch:2004pa},
are presented in Mellin space as a weighted sum of harmonic sums.
The latter is presented in $x$ space as a weighted sum of harmonic polylogarithms.
Our approach for determining the Mellin transform of these harmonic polylogarithms is discussed in Appendix \ref{NNLOHS}.

Because the effect of the observed hadron's mass is expected to be significant at low $x$, we incorporate the hadron mass effects
according to the method of Ref.\ \cite{Albino:2005gd}.
In this case, the scaling variable $x$, which in the factorization theorem
is defined as the ratio of the detected hadron's light cone momentum to the overall
process's, must be distinguished from the energy and momentum fractions measured in experiment and
given by
\be
x_E = 2E_h / \sqrt s\qquad and \qquad x_p = 2\vert \overrightarrow{{\bf p}_h}\vert / \sqrt s
\ee
 respectively.
We stress that $x_E$ and $x_p$ equal $x$ only when hadron mass effects are neglected.
Otherwise, they are related to $x$ via:
\be
x_p = x \left(1-\frac{m_h^2}{s\,x^2}\right),\qquad x_E
= x \left(1+\frac{m_h^2}{s\,x^2}\right).\label{mh}
\ee
The cross sections $d\sigma^{K^\pm} /dx$ and $d\sigma^{K^0_s} /dx$ that determine $d\sigma^{\cal K}/dx$,
 eq. (\ref{basic}), which we are calculating and which enters the factorization theorem, are related to the measurable ones
$d\sigma /dx_p$ and $d\sigma /dx_E$ via~\cite{Albino:2005gd}:
\be
\frac{d\sigma}{dx_p}\,(x_p,s)&=&\frac{1}{1+m_h^2/[sx^2(x_p)]}\,\frac{d\sigma}{dx}(x(x_p),s)\\
\frac{d\sigma}{dx_E}\,(x_E,s)&=&\frac{1}{1-m_h^2/[sx^2(x_E)]}\,\frac{d\sigma}{dx}(x(x_E),s),
\ee
where $d\sigma $ stands for either $d\sigma^{K^\pm}$ or $d\sigma^{K^0_S}$.
We exploit the fact that different data groups use different definitions for ``$x$''
in order to obtain the kaon mass, by using the above relations and fitting the mass $m_h$.
We assume the masses of the neutral and charged kaons are equal.
We note that fitting

The total number of free parameters in our fits to charged and neutral kaon data is 66,
and the total number of data points is 730.

We could choose to fit the data at a subset of $\sqrt{s}$  values and then predict the remaining data using the universality
of the non singlet FF thus obtained. However, we choose to simultaneously fit all the available data in order to maximize
the constraints on the parameters appearing in eq.\ (\ref{1stkaonnonsing}).
As we will see in section \ref{results},
the simultaneous description of all data with the same fitted non singlet FF turns out to be good,
in accordance with the universality of FFs.

\section{Results of the analysis}
\label{results}

First we perform a fit
only to the available $K_S^0$ production data in order to ensure that the parametrization in eq.\ (\ref{paramofK0Sdata})
is adequate for the $K_S^0$ data
that we will use in our extractions of the kaon non singlet.
Then we perform a simultaneous fit
to both charged and neutral kaon production data in $e^+ e^-$ reactions.
In the latter case we perform our analysis to NLO and NNLO in perturbative QCD.

\subsection{Analysis of $K^0_S$ data \label{KSonly}}

Here we present our results from a fit to $K^0_S$ data only, using the parametrizations in eq.\
(\ref{paramofK0Sdata}). The average $\chi^2$ per data point, $\chi^2_{\rm DF}$,
for each data set is presented in Table \ref{K0Stable} together with details of the data set.
Also shown, where applicable, is the value at the global minimum of $\lambda$ for each data set,
which after multiplication by the normalization error
is the most likely systematic deviation of the central values (see Ref.\ \cite{Albino:2008fy}
for a complete discussion), and which
should obey $|\lambda|\lesssim 1$ for a reasonable fit.

\begin{table}[h!]
\caption{Summary of the measurements for inclusive single $K_S^0$ production in $e^+ e^-$ reactions.
The column labeled ``Cross section'' gives the type of cross section measured,
up to the normalization and possible non zero width $x$ bins.
The column labeled ``\# data'' gives the number of data.
The column labeled ``Norm.\ (\%)'' gives the normalization uncertainty on the data as a percentage.
The values of $\lambda$ and $\chi^2_{\rm DF}$ from the fit described in the text are also given.
In this fit the fitted mass is $m_K=320$ MeV.
\label{K0Stable}}
\begin{center}
\begin{tabular}{|c|c|c|c|c|c|c|c|}
\hline
\multirow{2}{*}{Collaboration} & \multirow{2}{*}{Cross section} & \multirow{2}{*}{Tagging}
& \multirow{2}{*}{\vspace{0.3cm} $\sqrt{s}$} & \multirow{2}{*}{\vspace{0.3cm} \#}
& \multirow{2}{*}{\vspace{0.3cm} Norm.\ } & \multirow{2}{*}{$\chi^2_{\rm DF}$} & \multirow{2}{*}{$\lambda$}\\
& & & (GeV) & data & (\%) & & \\
\hline
\input{K0Stable}
\hline
\end{tabular}
\end{center}
\end{table}

In general, as seen from the Table, our parametrization provides a good description
of all but the HRS data, where the description is poor.
At $\sqrt{s}=91.2$ GeV, the $b$ quark tagged data appears to be slightly inconsistent with the other data.
The value of $|\lambda|$ for the ALEPH data is high, but the fit to the other data at $\sqrt{s}=91.2$ GeV in general is good.
Otherwise, both $\chi^2_{\rm DF}$ and $|\lambda|\lesssim 1$ which suggests that the parametrization
in eq.\ (\ref{paramofK0Sdata}) is sufficient to represent these data.

In the caption of Table \ref{K0Stable} we quote the fitted kaon mass $m_K=320$ MeV, which
is somewhat smaller than the true mass of 498 MeV. However, it is not significantly different from
the value 343 MeV obtained in global fit analyses in
Ref.\ \cite{Albino:2008fy}, where it was argued that kaon production through
complex decay chains may cause a significant difference
between the true mass and the fitted mass, when only direct parton fragmentation is assumed in the calculations.

\subsection{Analysis of $K_S^0$ and $K^\pm$ \label{mainfit}}

Here we present the results from the combined analysis of $K^\pm$ and $K^0_S$ data.

We implement large $x$ resummation in our NLO analysis.
We resum both leading and next-to-leading logarithms (LL and NLL respectively), which are all the classes of
logarithms appearing at this order.
As shown in the AKK08 fit \cite{Albino:2008fy}, this
significantly improves fits to charged kaon data at large $x$.
Resummation in the quark cross section (or quark coefficient function)
is obtained from the method of Ref.\ \cite{Albino:2000cp} and the results
for the unfactorized partonic cross section in Ref.\ \cite{Cacciari:2001cw},
while resummation in the evolution is performed according to the method in Ref.\ \cite{Albino:2007ns}.

Thus we apply the two most optimum theoretical tools to our calculations, namely the NLO results with resummation,
and the NNLO without resummation.

The measured inclusive $K^\pm$ and $K^0_S$ production cross sections and the obtained $\chi^2_{\rm DF}$
values, both in NLO and NNLO, are shown in Table \ref{Ktable}.
\begin{table}[h!]
\caption{As in Table \ref{K0Stable}, but for the fit to both $K^\pm$ and $K^0_S$ production data,
in which the perturbative components in the cross section differences $d\sigma^{K^\pm}-2d\sigma^{K_S^0}$
are calculated in NLO and NNLO. The fitted mass of the kaon was $m_K({\rm NLO})=$124 MeV and
$m_K({\rm NNLO})=$55 MeV. \label{Ktable}}
\begin{center}
\begin{tabular}{|c|c|c|c|c|c|c|c|c|c|c|}
\hline
\multirow{2}{*}{Collaboration} & \multirow{2}{*}{Cross section} & \multirow{2}{*}{Tagging}
& \multirow{2}{*}{\vspace{0.3cm} $\sqrt{s}$} & \multirow{2}{*}{\vspace{0.3cm} \#}
& \multirow{2}{*}{\vspace{0.3cm} Norm.\ } & \multirow{2}{*}{\vspace{0.3cm}$\chi^2_{\rm DF}$} & \multirow{2}{*}{$\lambda_{\rm NLO}$}
& \multirow{2}{*}{\vspace{0.3cm}$\chi^2_{\rm DF}$} & \multirow{2}{*}{$\lambda_{\rm NNLO}$}\\
& & & (GeV) & data & (\%) & NLO& &NNLO&\\
\hline %\hline
\input{Ktable}
\hline
\end{tabular}
\end{center}
\end{table}
In general, with the exception of a few data sets, in particular the $b$ quark tagged cross section measurements,
the description of the data is rather good.
However, the kaon mass, both in NLO and NNLO,
is significantly lower than the one obtained in the phenomenological description
of the $K^0_S$ data (see Table \ref{K0Stable}) only, i.e.\ without perturbative QCD, also
it is significantly lower than the value 343 MeV obtained in Ref.\ \cite{Albino:2008fy}.

\begin{figure}[h!]
\begin{center}
\parbox{.49\linewidth}{\includegraphics[angle=-90,width=9.5cm]{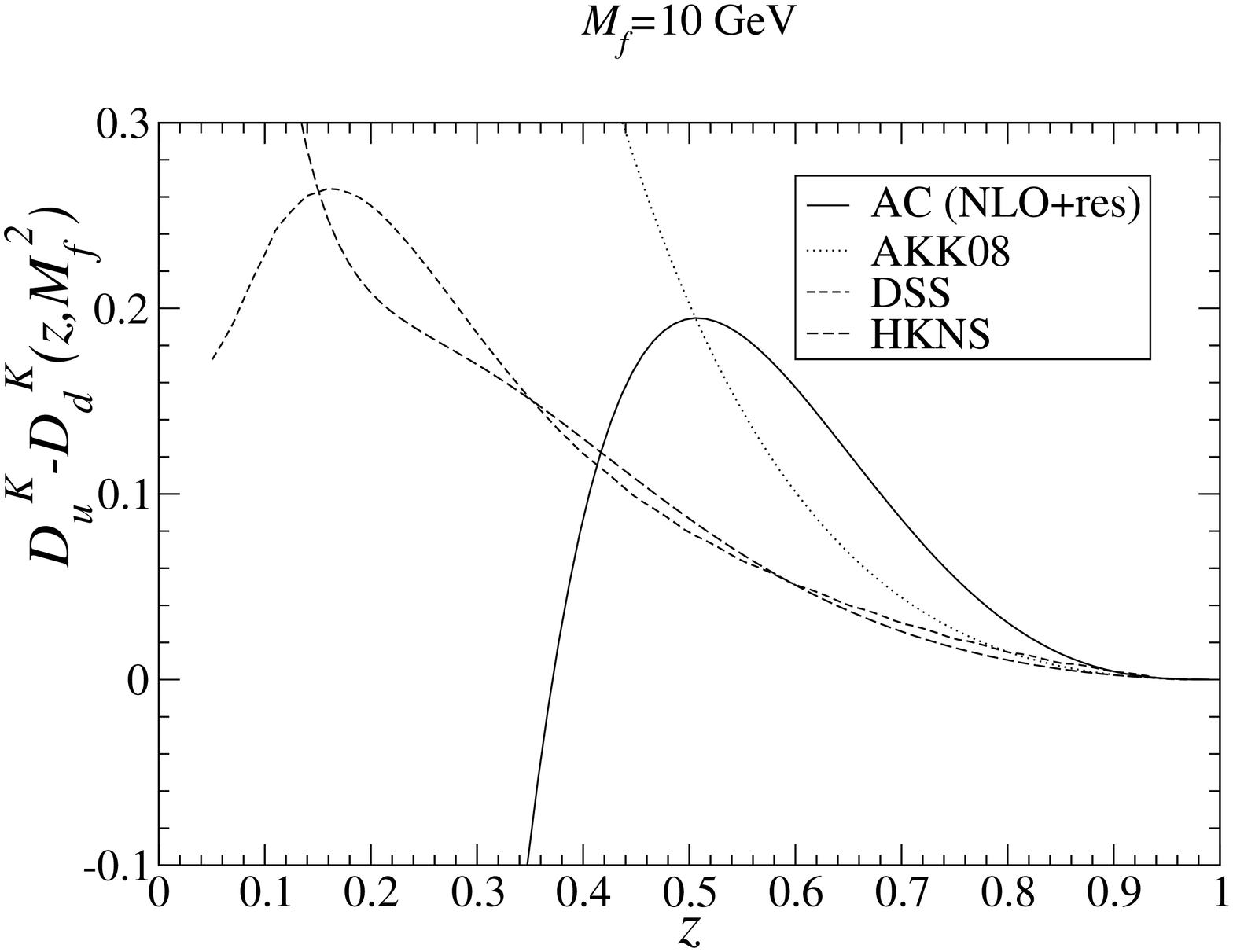}}
\parbox{.49\linewidth}{\includegraphics[angle=-90,width=9.5cm]{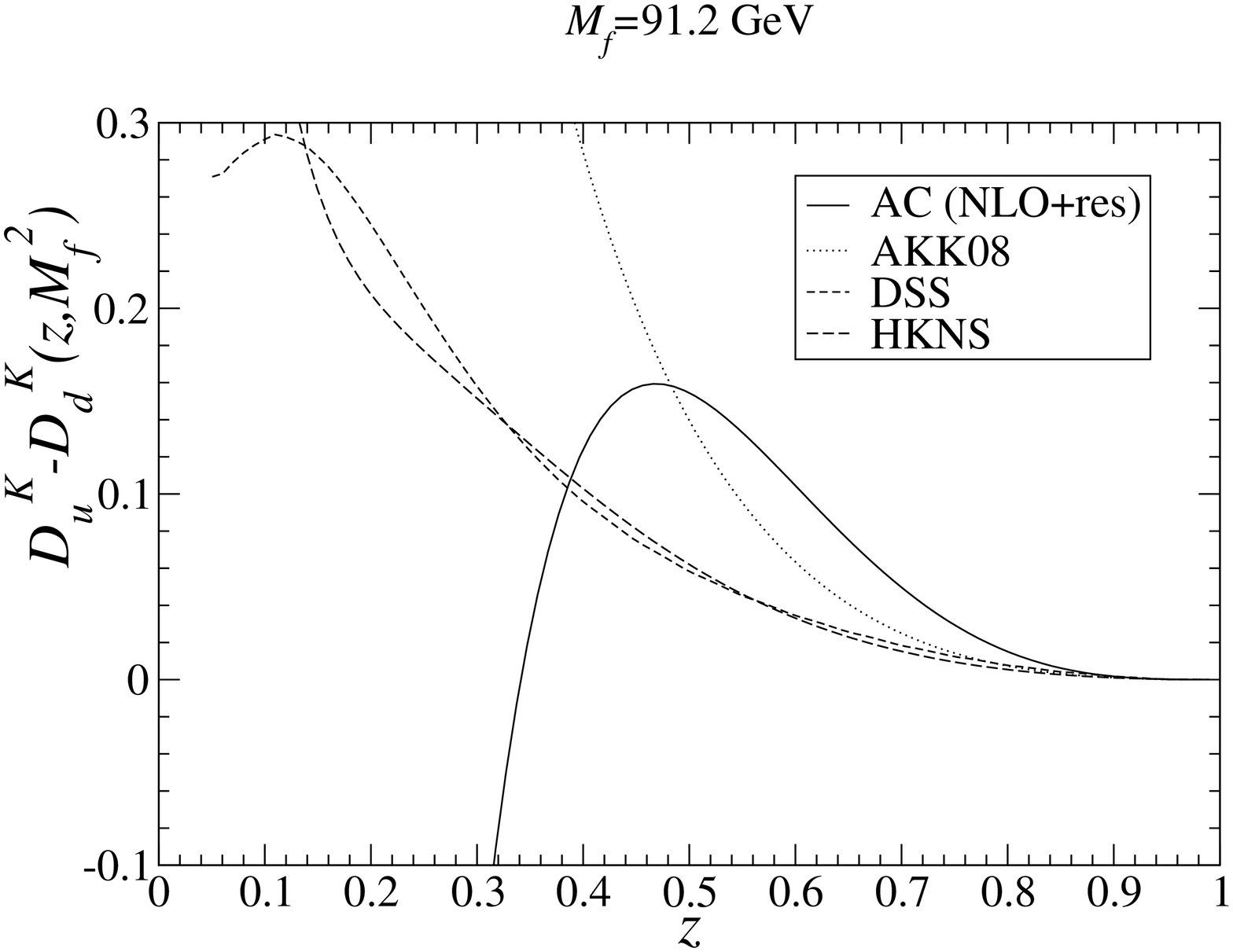}}
\caption{The kaon non singlet FF obtained in this paper at NLO with large $x$ resummation (labeled ``AC (NLO+res)'')
and from the calculations of the same quantity from the HKNS \cite{Hirai:2007cx},
DSS \cite{de Florian:2007hc} and AKK08 \cite{Albino:2008fy} FF sets.
\label{figcompnsff}}
\end{center}
\end{figure}

Our results for $D_u^{K^\pm}-D_d^{K^\pm}$ in NLO are shown in Fig.\ \ref{figcompnsff}. In the same figure
the NLO results from global fits of the DSS, HKNS and AKK08 sets are presented as well.
As seen from the figure, at $z\gtrsim 0.5$ there is an agreement in shape
among the different plots of the NS, but our NS is in general larger in magnitude. However, they
differ significantly at $z \lesssim 0.5$. The most striking difference is the negative value
for the NS at $z \lesssim 0.4$ obtained in our approach,
while all global fit parametrizations imply a positive $D_u^{K^\pm}-D_d^{K^\pm}>0$.

In Tables \ref{partable1} --- \ref{partable3} we show the values of the parameters for our main fit,  NLO + resummation.
However, we caution the reader that, because we begin our evolution at $\mu_f=\sqrt{2}$ GeV,
 due to neglect of higher order NNLO terms,  the uncertainties on these parameters in the NLO calculation may be very large
and thus may depend significantly on the method used for solving the DGLAP equations.
This uncertainty is approximately equal to the size of the NNLO terms.

\begin{table}[htb]
%\begin{center}
%\begin{minipage}{0.31\textwidth}
\caption{The fitted values of the parameters for
$(D_u^{K^\pm}-D_d^{K^\pm})(z,\mu_{f0}^2)$ parametrized as in eq.\ (\ref{1stkaonnonsing}),
from our main fit.}
\label{partable1}
\centering
\begin{tabular}{|c|c|}
\hline
Parameter & Value \\
\hline \hline
$n$ & -6.25\\
\hline
$a$ & -0.11\\
\hline
$b$ & 3.12\\
\hline
$n'$ & 11.13\\
\hline
$a'$ & 0.60\\
\hline
$b'$ & 3.01\\
\hline
\end{tabular}
%\end{minipage}
\end{table}

\begin{table}[htb]
\begin{center}
\caption{The fitted values of the parameters for
$d\sigma^{K_S^0}_{e^+e^-}/dx(x,s)$ parametrized as in eq.\ (\ref{paramofK0Sdata})
in the different energy intervals $\sqrt s$
 from our main fit: NLO with resummation.}
\label{partable2}
\begin{tabular}{|c|c|c|c|c|c|c|c|}
  \hline
  energy interval in [GeV] & N & $\Delta N$ & A & B & c & d & e \\
  \hline
  12 $< \sqrt s < $ 14.8 &  $1.58\times 10^{-5}$ &  $7.90\times 10^{-5}$ &  -17.2 &  -2.33 &  10.1 &  -2.99 &  -0.357 \\
  21.5$< \sqrt s < $ 22  & $8.43\times 10^5$  &  $-4.79\times 10^5$ &  14.5 &  12.2 &  -7.24 &  1.45 &  $8.81\times 10^{-2}$ \\
  29 $< \sqrt s < $ 35  &  -0.444 &  3.81 &  -2.57 &  3.49 &  1.79 &  -0.696 &  $9.25\times 10^{-2}$ \\
  42.6 $< \sqrt s < $ 44  &  $3.55\times 10^4$ &  $-1.32\times 10^5$ &  4.91 &  7.10 &  -1.66 &  $4.08\times 10^5$ & $-3.21\times 10^{-2}$  \\
  $\sqrt s$ = 58  & 6.03  &  0 (fixed) &  -1.23 &  6.26 &  1.11 &  -0.415 &  $-4.93\times 10^{-2}$ \\
  $\sqrt s$ = 91.2 &  $16.1$ &  0 (fixed) &  -4.88 &  -0.681 &  1.68 &  -0.338 &  $-2.92\times 10^{-2}$ \\
  183 $< \sqrt s < $ 189  &  $126.$ &  -646. & -1.34  &  5.16 &  0.492 &  -0.156 &  $-1.69\times 10^{-2}$ \\
  \hline
\end{tabular}
\end{center}
\end{table}

\begin{table}[htb]
\begin{center}
\caption{The fitted values of the parameters for
$d\sigma^{K_S^0}_{e^+e^-}/dx(x,s)$
parametrized as in eq.\ (\ref{paramofK0Sdata}) for $\sqrt s = 91.2$ GeV
 from our main fit: NLO with resummation for the $c$ and $b$ tagged data}
\label{partable3}
\begin{tabular}{|c|c|c|c|c|c|c|c|}
  \hline
  the data & N & $\Delta N$ & A & B & c & d & e \\
  \hline
  $c$ tagged &  2.62 &  0 (fixed) &  -7.52 &  2.39 &  3.97 & -1.01  &  $-9.35\times 10^{-2}$ \\
  $b$-tagged &  0.164 &  0 (fixed) &  -8.65 &  2.01 &  2.95 &  -0.474 &  $-3.08\times 10^{-2}$ \\
  \hline
\end{tabular}
\end{center}
\end{table}

\begin{figure}[h!]
\begin{center}
\parbox{.49\linewidth}{\hspace{1cm}\includegraphics[angle=-90,width=9.5cm]{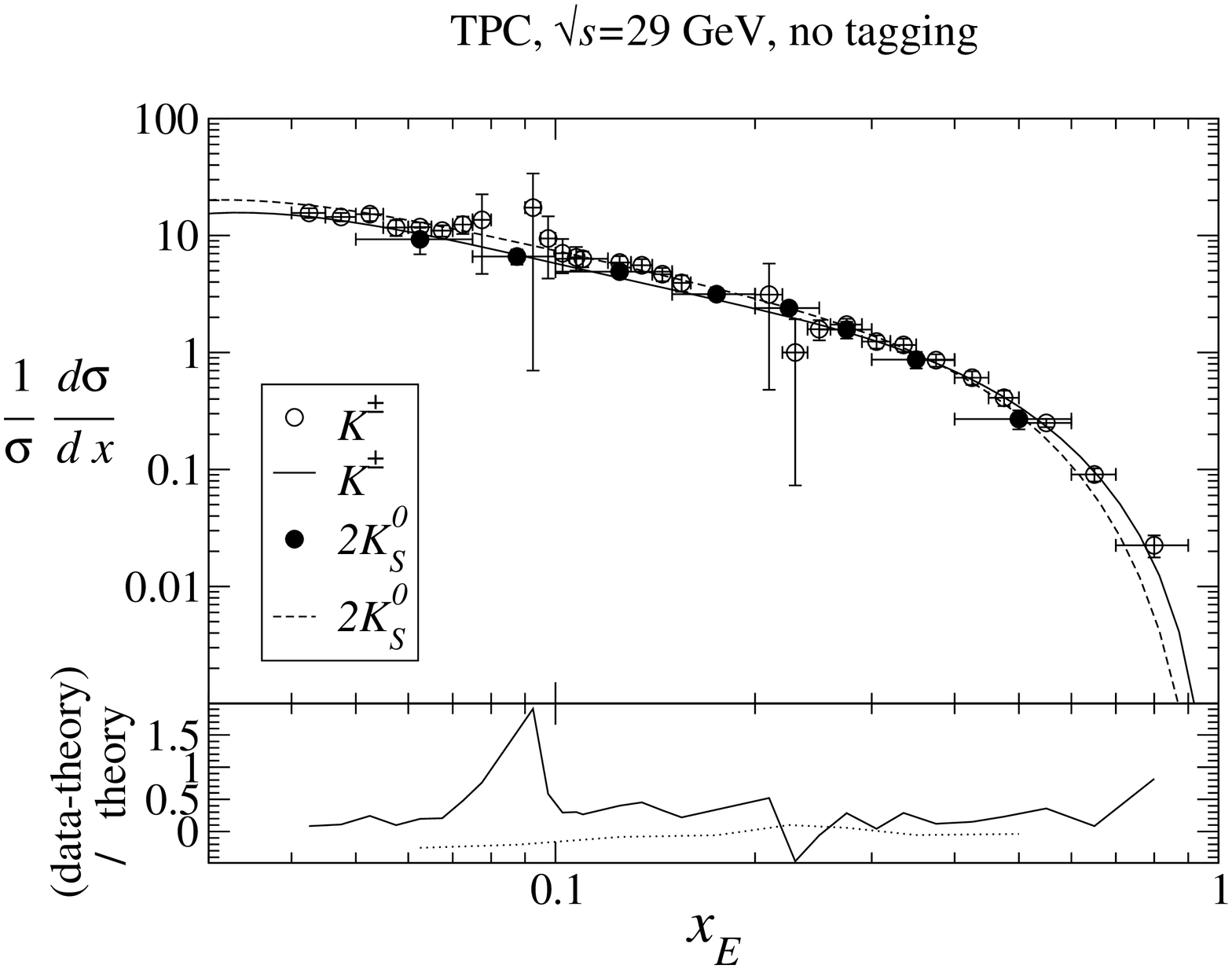}}
\parbox{.49\linewidth}{\hspace{1cm}\includegraphics[angle=-90,width=9.5cm]{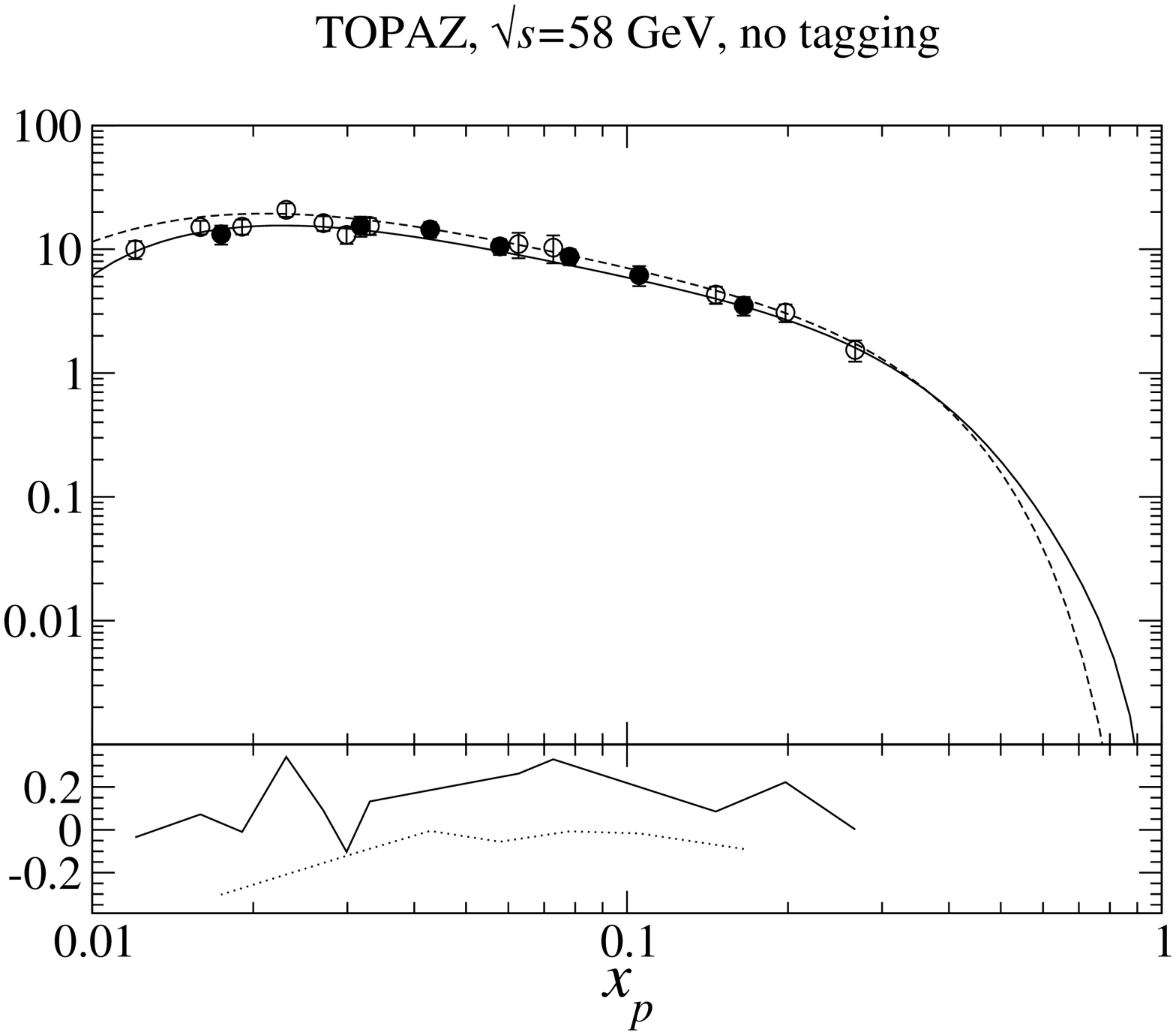}}\\[-0.2cm]
\parbox{.49\linewidth}{\hspace{1cm}\includegraphics[angle=-90,width=9.5cm]{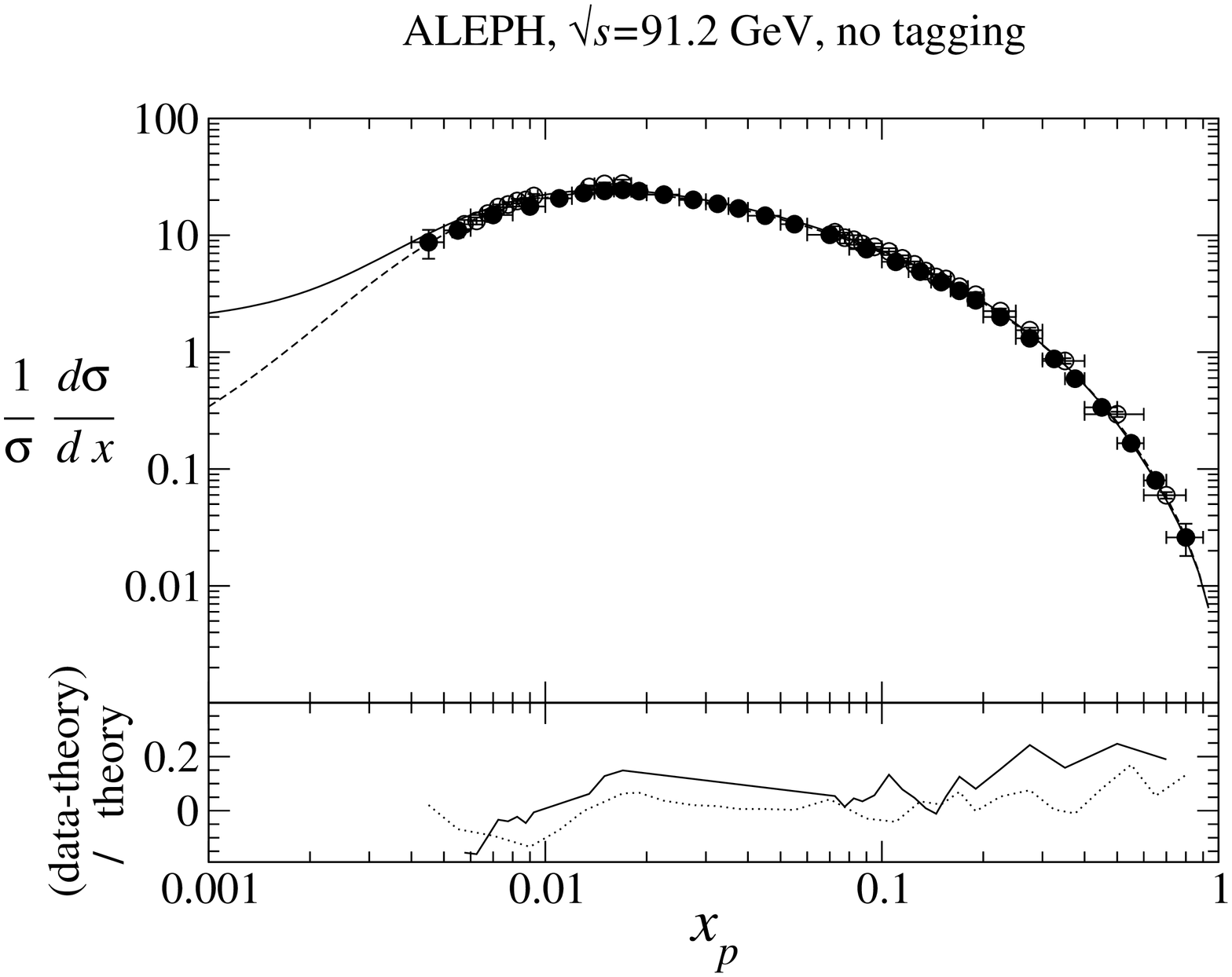}}
\parbox{.49\linewidth}{\hspace{1cm}\includegraphics[angle=-90,width=9.5cm]{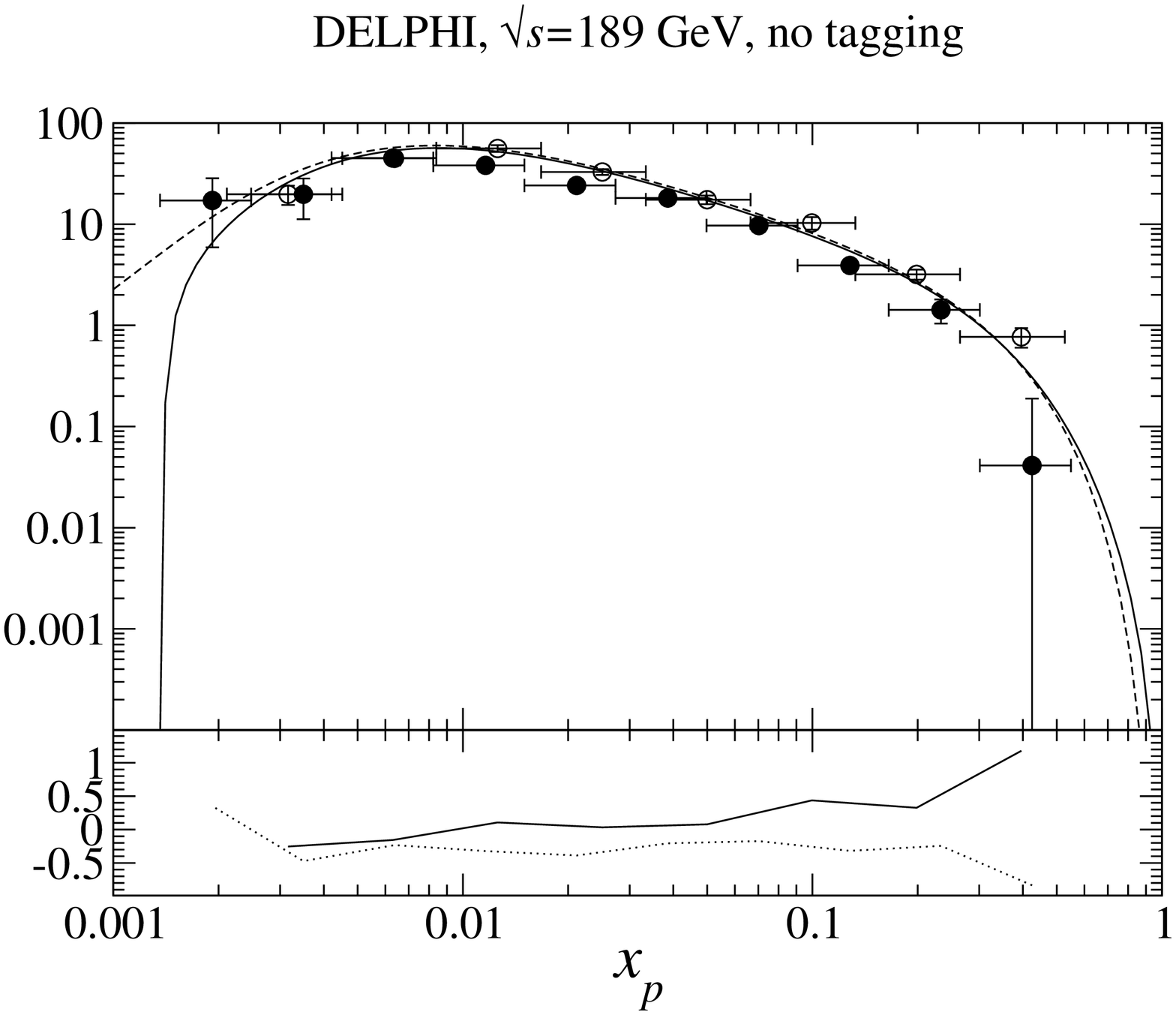}}
\caption{Comparison of charged and neutral kaon production at various c.m.\ energies.\label{figcompcandn}}
\end{center}
\end{figure}

In order to understand the origin of the negative value of
$D_u^{K^\pm}-D_d^{K^\pm}$ obtained from the difference cross sections
$\sigma^{\cal K}=\sigma^{K^\pm}-2\sigma^{K^0_s}$, we make a
comparison of the charged and neutral kaon production data
at various $\sqrt{s}$ in Fig.\ \ref{figcompcandn}.
Such a direct comparison is possible because, for these data, the
cross section measurements happen to be defined the same way, i.e.\
they are differential in the same variable and normalized in the
same way, which is not typical for the data in general.
In general, the description of these data is good.
According to the (data-theory)/theory plots,
the calculation for the $K_S^0$ production data tends to overshoot the central values of the data,
while for $K^\pm$ production the behaviour is the opposite, but this is not
significant relative to the experimental errors.
For $x\gtrsim 0.3$, the calculated charged kaon production exceeds the
neutral except when $\sqrt{s}=91.2$ GeV. However, below this region
in $x$ the opposite behaviour is observed,
i.e.\ $\sigma^{\cal K} <0$ for
$x\lesssim 0.3$, for all $\sqrt s$ except $\sqrt{s}\simeq 91.2$ GeV, where $\sigma^{K^\pm}$ and
$\sigma^{K^0_s}$ are very similar. As Fig.\ \ref{figeued} shows,
the sign of the difference of the effective electroweak couplings of the
$u$ and $d$ quark flavours, for all $s$ except around the $Z$-pole
($78<s<122$ GeV$^2$), is positive, i.e.\ $\hat e_u^2-\hat e_d^2 >0$
for $s\,\gtrsim\, 78$ GeV$^2$ and $s\,\lesssim\,112$ GeV$^2$, i.e. at
the cross sections that give the main contribution to the NS in
(\ref{basic}). Then, following the simple LO approach in which
convolutions are replaced by ordinary products, eq.(\ref{basic})
implies that $D_u^{K^\pm}-D_d^{K^\pm} <0$ at $z\lesssim 0.3$, and
$D_u^{K^\pm}-D_d^{K^\pm} >0$ at $z\gtrsim 0.3$. Of course these rough
arguments do not take into account experimental errors, which are
rather big for kaon production, or convolutions etc., however they do
help to verify the result qualitatively.

Our negative result for $D_u^{K^\pm}-D_d^{K^\pm}$ at low $z$,
though justified by the above arguments on the data on
$\sigma^{\cal K}$, is however in contrast to the intuitive
interpretation for favoured $u$-quark and unfavoured $d$-quark
transitions. In addition, our result is quite different from the
DSS, AKK08 and HKNS results. There could be several reasons for
this, as well as for the unexpectedly low values for the kaon
mass $m_K({\rm NLO})=124$ MeV and $m_K({\rm NNLO})=55$ MeV, shown in
Table \ref{Ktable}.
Most probably it is due to the different assumptions in the parametrizations
and to inclusion of the small $x$-data in our fit. The DSS and HNKS
collaborations use the assumption that all light quark unfavoured FFs are
equal: $D_{\bar u}^{K^+}=D_s^{K^+}=D_{d}^{K^+}=D_{\bar d}^{K^+}$,
while no assumptions were used in the AKK08 fit and in the analysis
in this paper, denoted by AC.
The fact that the DSS and HKNS non singlet FF, which can be written as $D_u^{K^+}+D_{\bar u}^{K^+}-D_d^{K^+}-D_{\bar d}^{K^+}$,
is lower than the others for $z\gtrsim 0.4$ in Fig.\ \ref{figcompnsff}
suggests that $D_d^{K^+}$ and $D_{\bar d}^{K^+}$ may be overestimated in this region when the light quark unfavoured FFs
are fixed to be equal to one another.
Maybe this could explain the similarity of the results for
$D_u^{K^\pm}-D_d^{K^\pm}$ obtained from the DSS and HKNS fits on one hand, and of AKK08 and AC at
$z\gtrsim 0.5$ on the other hand (see Fig.\ \ref{figcompnsff}).
The AKK08 and HKNS analyses used no data below
$x\leq 0.05$, the DSS analysis used only data at $x\geq 0.1$, while
we include data as low as $x\simeq 0.001$. The discrepancy may also
be a result of various low $x$ effects not accounted for in the
calculation, such as dynamical higher twist, quark mass corrections,
etc. However, perhaps the most likely reason are the large
experimental errors on the NS.
The FFs of the various collaborations should be the same within the error
(composed of the theoretical errors and the (unknown) experimental
errors propagated from the fitted data to the FF).
Thus if we assume that the various FFs are consistent, then the spread of FFs in Fig.\ \ref{figcompnsff}
gives some indication of the error on the FF, and shows the error increasing drastically with decreasing $z$.
This argument assumes that the (similar) assumptions made on the FFs in the DSS and HKNS fits are correct.
In any case, these results warrant
further investigation into the validity of the standard approach at
low $x$. It is promising, however, that it is possible to fit low
$x$ data ($x\simeq 0.001$) using fixed order perturbation theory.

The  negative value of the non singlet FF $D_u^{K^+}+D_{\bar u}^{K^+}-D_d^{K^+}-D_{\bar d}^{K^+}$
at low $z$ contradicts the physical argument that favoured FFs are larger than unfavoured FFs.
This behaviour alone is not too serious since FFs in general are not physical.
However, the second Mellin moment of a FF $D_a^h$ is physical,
in the sense of being factorization scheme and scale  independent, and can be
interpreted as the fraction of momentum of the fragmenting parton $a$ that is carried away by hadrons of species $h$.
The second moment of the non singlet FF is expected then to be positive, but from our fit using calculations to NLO (NNLO)
the result is -0.07 (-0.1). Most likely this is a consequence of the large experimental errors at low $z$.
However, it could also arise from a breakdown of perturbation theory, or from effects not accounted for in
 the calculation at low $x$, if such effects turn out  significantly large.

In order to check our negative result for the NS at small $z$, we performed
a fit in which a parameterization of the form $n z^a (1-z)^b$,
instead of that in eq.\ (\ref{1stkaonnonsing}), was used which
yielded a positive NS (i.e.\ $n>0$).
However, the result $\chi^2_{\rm DF}=2.4$ was obtained,
which corresponded to a $\chi^2$ of about 100 points above that for our main fit.
Thus the parametriztion in eq.\ (\ref{1stkaonnonsing}) is much more favoured by the data.
We also performed a fit in which the NS was fixed to zero, and obtained $\chi^2_{\rm DF}=2.4$ again.
Thus a positive, as well as a zero kaon non singlets are both allowed by data as a whole, but the fits
are much worse. Note that in our analysis we include data at very low $x$,
which are the most accurate data and any deviations
of the fit from these data immediately results in higher $\chi^2$. It is the small $x$ data that raises $\chi^2$ with
the zero and $n z^a (1-z)^b$ parameterizations.

\begin{figure}[h!]
\begin{center}
\parbox{.49\linewidth}{\includegraphics[angle=-90,width=9.5cm]{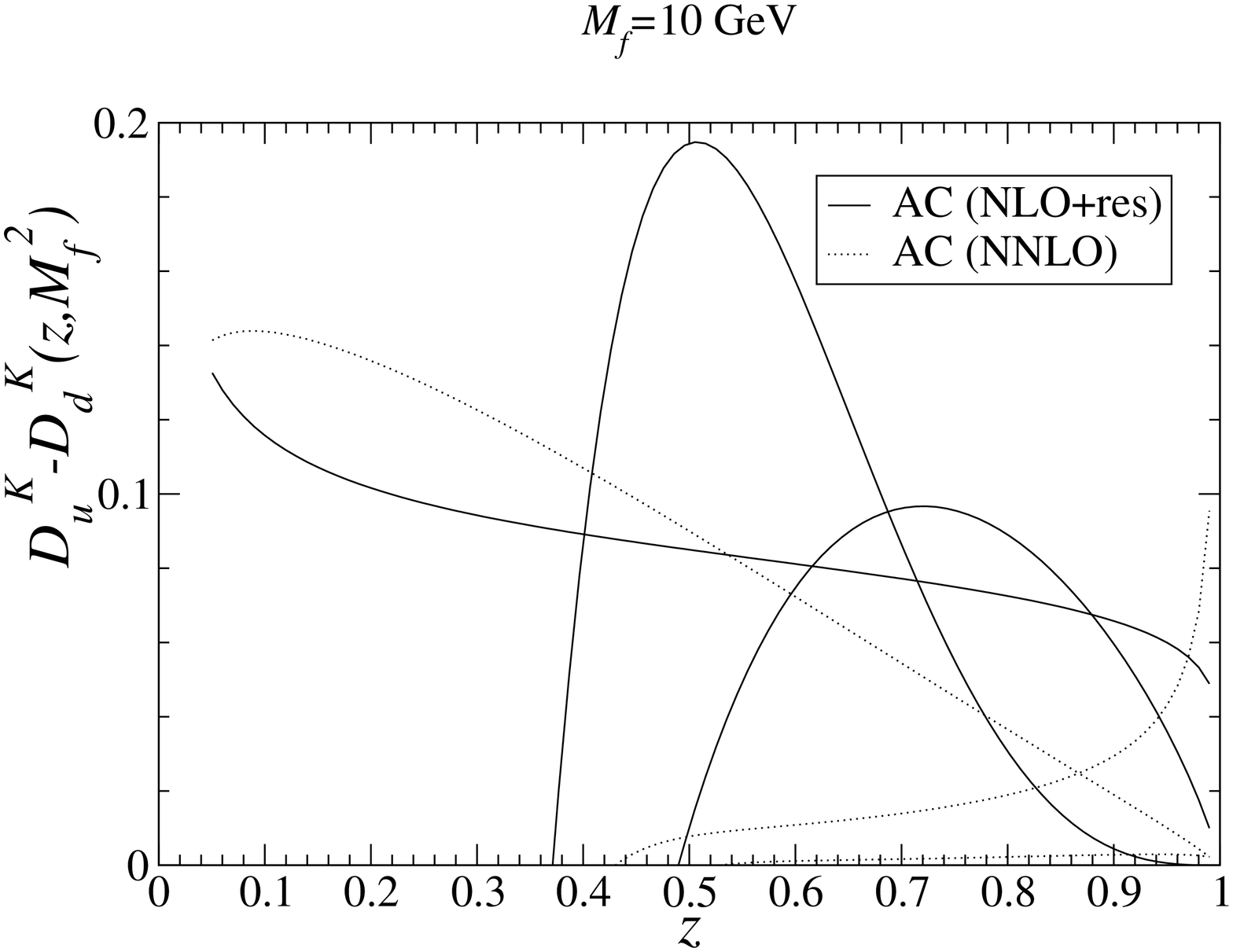}}
\parbox{.49\linewidth}{\includegraphics[angle=-90,width=9.5cm]{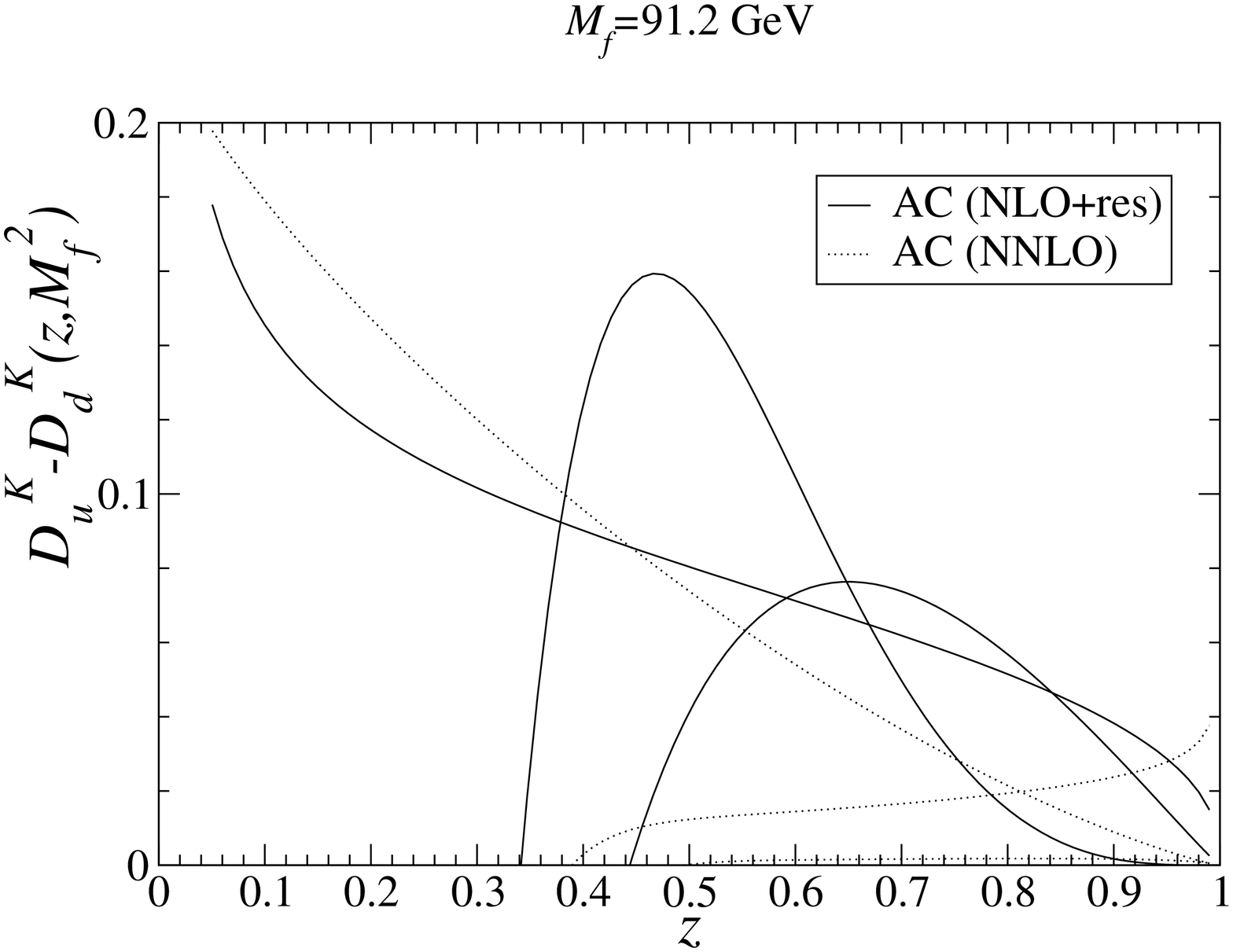}}
\caption{The kaon non singlet FF obtained in this paper at NLO with large $x$-resummation and at NNLO without resummation
from fits for which $k=\mu_f^2/s=\mu^2/s=$1/4, 1 and 4.
For both the left and right plots, the fits for the NS FF
are arranged in their increasing magnitude at $z=0.8$ as follows:
at NNLO with $k=1$ and then with $k=1/4$, at NLO with $k=1$, at NNLO with $k=4$
and at NLO with $k=4$ and finally with $k=1/4$.
Note that the fit at NNLO with $k=1$
yields almost a zero NS, i.e.\ the curve coincides with the $z$-axis.
\label{figcompnsffscale}}
\end{center}
\end{figure}

In Fig.\ \ref{figcompnsffscale} we compare the NLO and NNLO fits for
$D_u^{K^\pm}-D_d^{K^\pm}$ at $\mu_f=10$ GeV (Fig.\
\ref{figcompnsffscale}, left) and at $\mu_f=91.2$ GeV (Fig.\
\ref{figcompnsffscale}, right) for various choices of the
factorization scale $k=\mu_f^2/ s$, $k=1,\,1/4,\,4$. As seen from
this figure,  there is an extremely strong dependence on the choice
of $k$ and on the chosen perturbative order -- NLO or NNLO. The
quality of the fits for all curves is good, which indicates that the
errors of the available data are too big to constrain the
non-singlet FFs. With very accurate data, the spread of the three
NLO FFs
 and the spread of the three NNLO FFs  should be less
and, assuming that the theoretical error in the calculations is
sufficiently larger than the experimental error,
 both should give some indication of the theoretical error.
 However, since large x resummation has been applied only to the NLO calculation and not in NNLO,
  this analysis  cannot
  test the perturbative convergence by comparing NLO and NNLO calculations. Also the low $x$-data, for the first time included
in an analysis, might have caused troubles. Fig.\
\ref{figcompnsffscale} implies only upper and lower bounds on the
non-singlet: 0 $\leq D_{u-d}^{K^\pm}\leq $ 0.2 at $z\geq 0.35$.

Note that in the used method,
the uncertainties of the FFs are due almost completely to the
experimental errors on the data. The theoretical error in the
calculation is relatively negligible here. Having the  experimental errors as they are, one
would not get such good fits to the NS FF from cross section (not
cross section difference) measurements, even with similarly large
errors. One of the reasons is that,  due to the large $x$-logarithms
in the singlet / gluon evolution and in gluon coefficient functions,
the  small $x$-data cannot be included in the analysis.
 All curves are positive for $z\gtrsim 0.5$, and the theoretical error in this region
is arguably less for the NNLO fits. More
accurate data is needed to better determine the scale variation.

The non singlet FF from the NNLO fit with $k=1$ is close to zero for $z\gtrsim 0.5$,
suggesting that the cross section $\sigma^{\cal K}$ is too.
This is a consequence of the fact that the $K^\pm$ production data and the $K^0_S$ production data (multiplied by 2)
are very close (see Fig.\ \ref{figcompcandn}).
This behaviour is consistent with our finding above, that a good fit can be obtained with the non singlet FF
fixed to zero (i.e.\ all parameters in eq.\ (\ref{1stkaonnonsing}) fixed to zero so that only the
parameters in eq.\ (\ref{paramofK0Sdata}) are varied in the fit).

\begin{figure}[h!]
\begin{center}
\parbox{.49\linewidth}{\includegraphics[angle=-90,width=9.5cm]{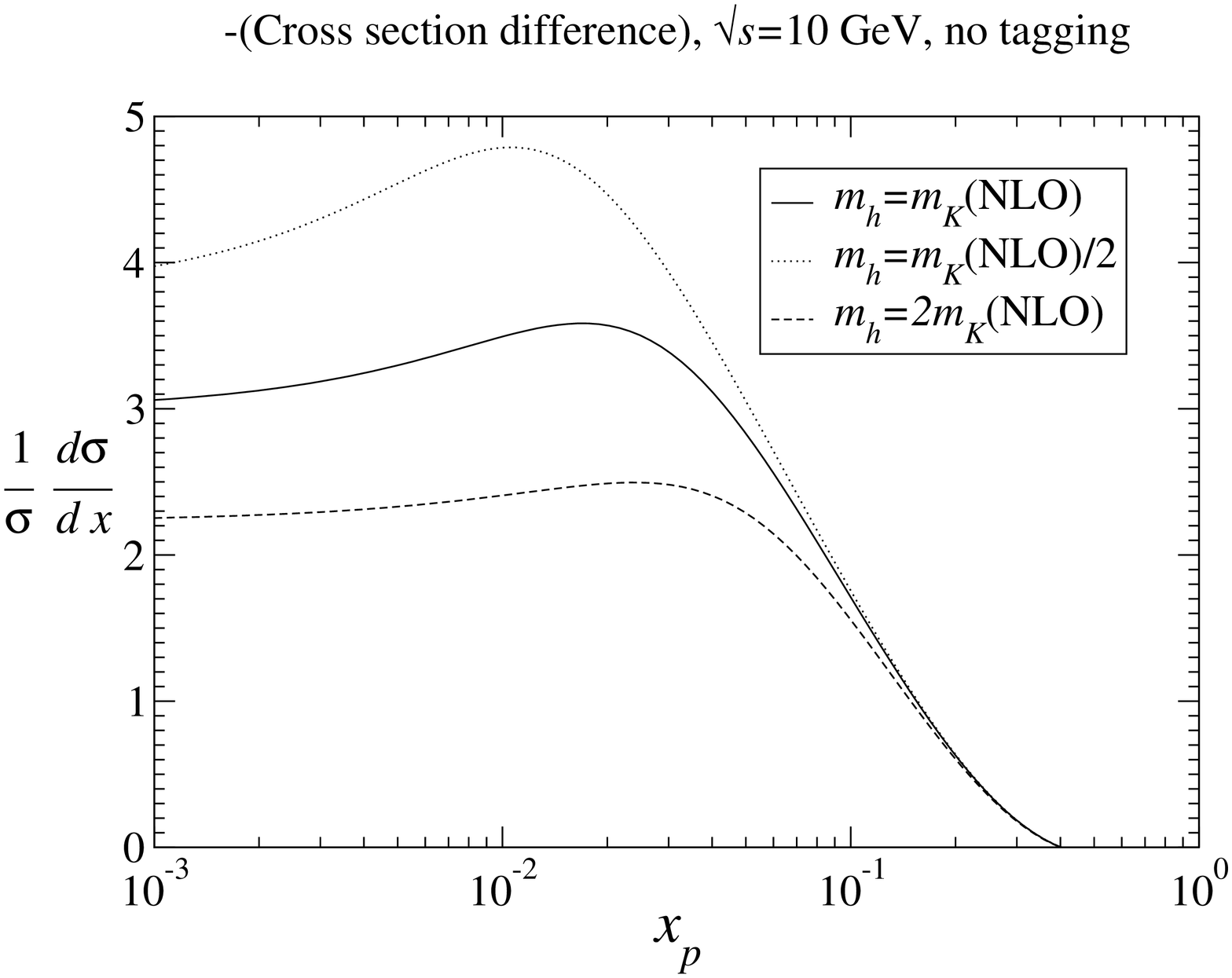}}
\parbox{.49\linewidth}{\includegraphics[angle=-90,width=9.5cm]{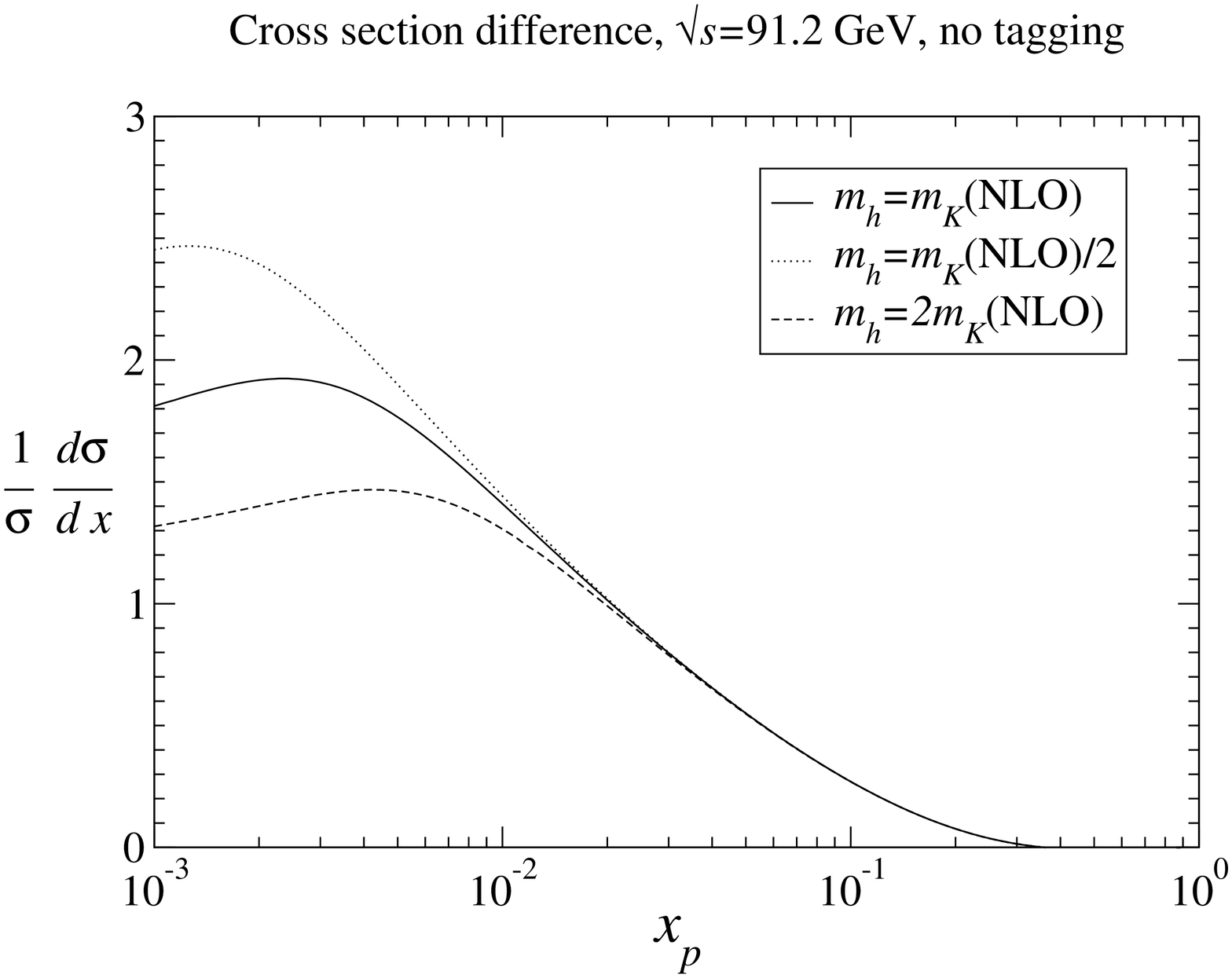}}
\caption{The fitted kaon cross section difference at different c.m.\ energies.
Also shown is the same quantity but with the kaon mass $m_h$ varied from its fitted result $m_K$.
Note that the left plot shows the negative cross section difference, $-\sigma^{\cal K}=2\sigma^{K^0_s}-\sigma^{K^\pm}$.
The curves in the left and right plot are negative (and not shown) for $x \gtrsim 0.4$ and 0.3 respectively.
\label{figcompmass2}}
\end{center}
\end{figure}

In Fig.\ \ref{figcompmass2}, we examine the sensitivity of the cross
section difference to the mass $m_h$ for two different values of $\sqrt s$, taking the
mass to be $m_h=m_K,\, m_K/2$, and $2\,m_K$ in eq.\ (\ref{mh}), where $m_K$ is the fitted mass.
As shown in the figure, for $\sqrt s$ = 10 GeV the calculation becomes sensitive to $m_h$
at $x\lesssim 0.1$,
while for $\sqrt s=91.2$ the sensitivity sets in at $x\lesssim 0.01$ where most of the data lie.
Because the calculation is very sensitive to hadron mass effects at low $\sqrt s$ and small $x$,
these effects strongly affects our fits.
Conversely, precisely because the hadron mass effects are important for the data in this region means that they cannot
be neglected.
However, other low $\sqrt s$, small $x$ effects, such as higher twist and mass effects of resonances from which
the kaon has been produced, will also be absorbed into $m_K$ after it has been fitted.
More accurate data will be needed to determine how important these other effects are.

\begin{figure}[h!]
\begin{center}
\parbox{.49\linewidth}{\includegraphics[angle=-90,width=9.5cm]{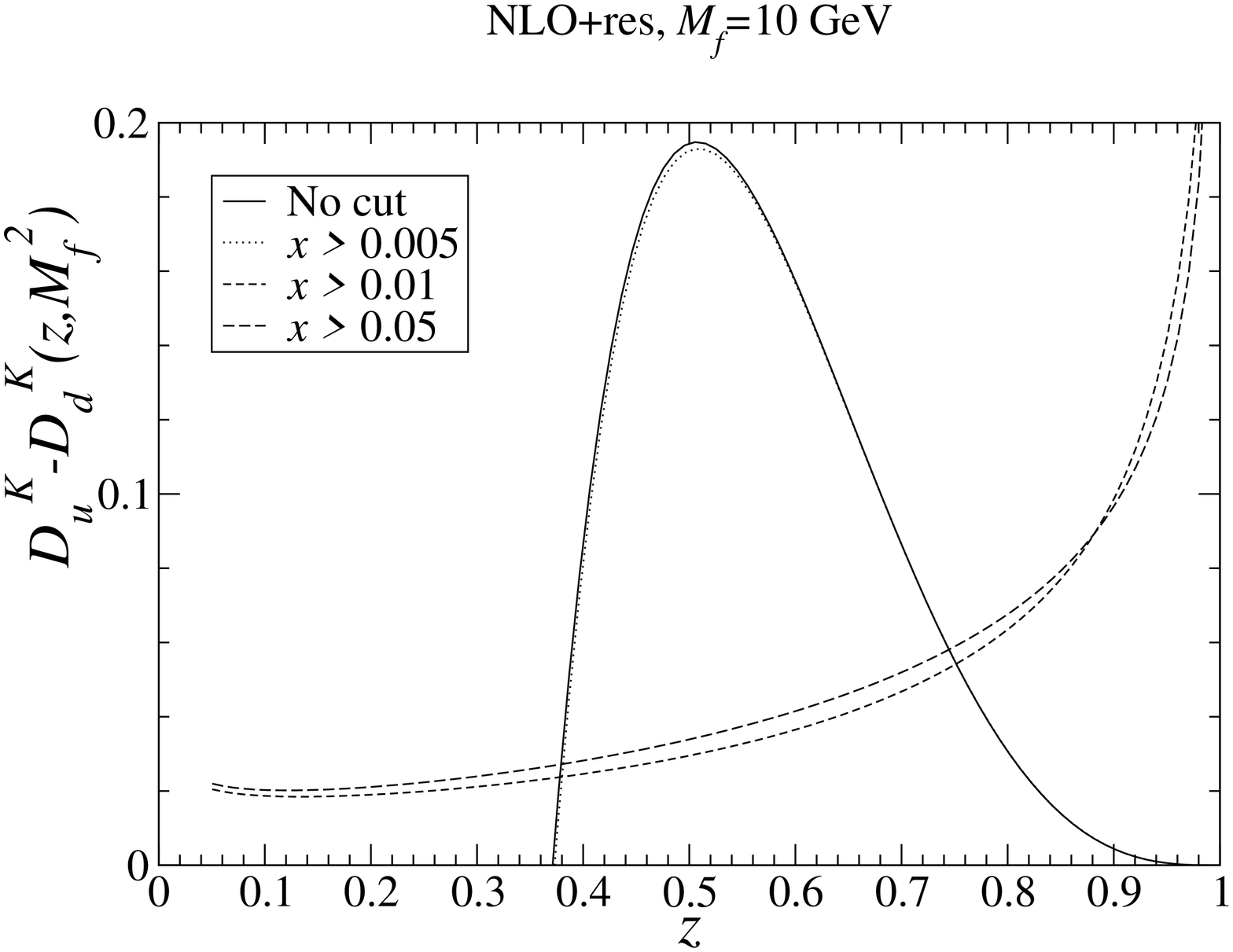}}
\parbox{.49\linewidth}{\includegraphics[angle=-90,width=9.5cm]{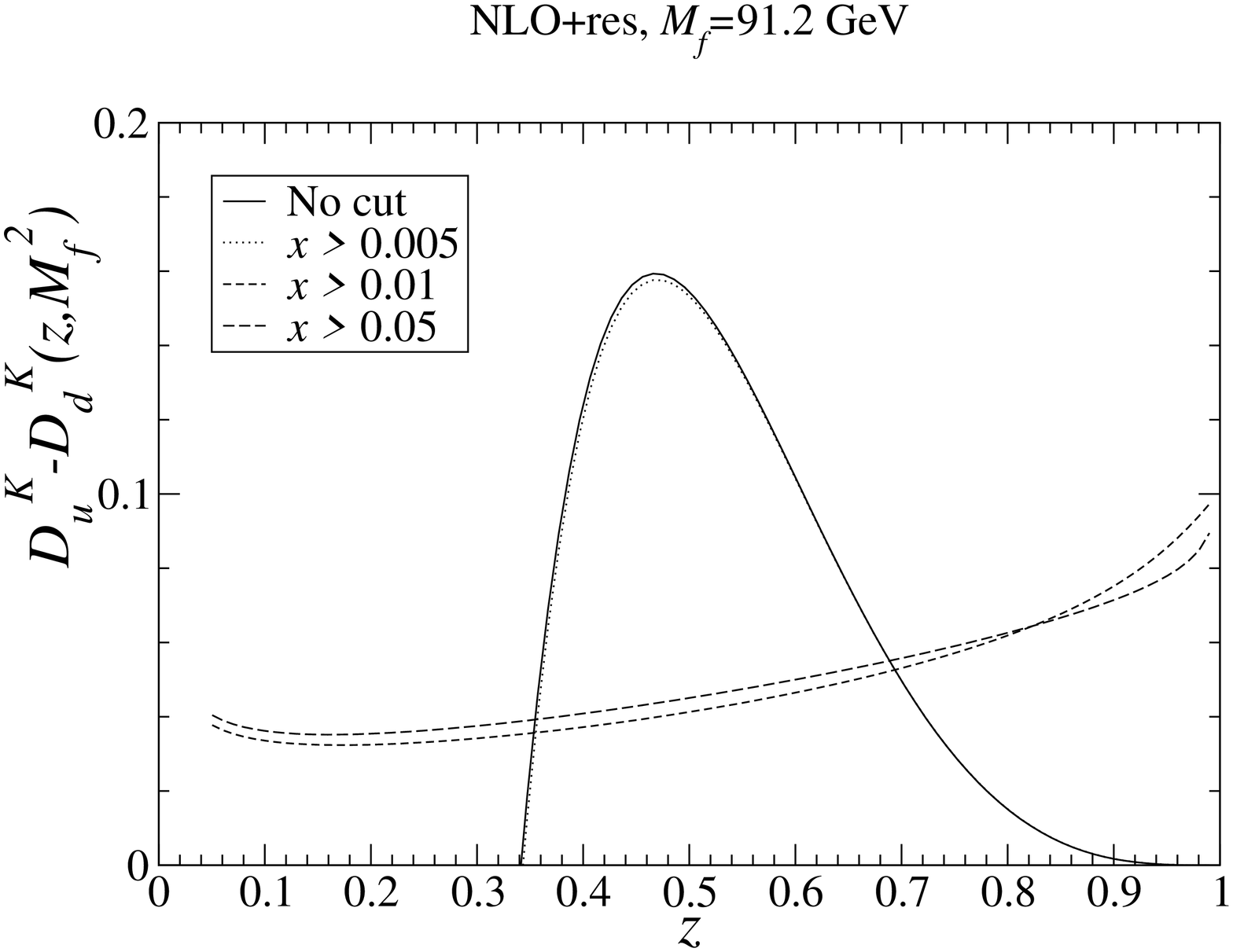}}
\caption{The kaon non singlet FF obtained in this paper at NLO
from fits for which various cuts on the data were imposed.
\label{figcompnsffcuts}}
\end{center}
\end{figure}

In Fig.\ \ref{figcompnsffcuts} we show the effect on the fitted NS of increasing the lower bound in $x$ on the data.
The result when no cut is imposed is similar to the result with a cut of $x>0.005$,
implying that data for which $x<0.005$ might not impose important constraints. The largest change in the fitted NS
is from $x>0.005$ to $x>0.01$.
The fact that this difference is so large suggests that a new (but valid) minimum in $\chi^2$ has been found.
This implies that the
accuracy of the data at $x > 0.1$ is not enough to form the difference cross
sections with the required precision to determine $D_u^{K^\pm}-D_d^{K^\pm}$.
In particular, note the unphysical divergence of the FF as $z\rightarrow 1$, which
is caused by the negative values of the fitted $b$ and $b'$ parameters in eq.\ (\ref{1stkaonnonsing}),
which further indicates the inability of the data at $x > 0.1$ to constrain the FF at large $z$.
Only including the large amount of precise small $x$ data, which through convolution determines
the $(D_u^{K^\pm}-D_d^{K^\pm})(z)$ not only at $z=x$, but also at all $z>x$, allows to determine
the NS in the whole $z$ region.

\section{Summary}
\label{summary}

The cross section difference $d\sigma^{\cal K}=d\sigma^{K^\pm}-2d\sigma^{K^0_S}$ determines uniquely the NS
$D_u^{K^\pm}-D_d^{K^\pm}$ without any assumptions. We have extracted $D_u^{K^\pm}-D_d^{K^\pm}$
from kaon production in $e^+ e^-\to K+X,\quad K=K^\pm,K^0_S$ and compared our results to those from global fit analyses,
namely the DSS, the HNKS and the AKK08 parametrizations.
In contrast to global fits, in our analysis i) data at much lower values of $x$, as low as $\simeq 0.001$,
could be included in the fit because
of the absence of SGLs in NS perturbative quantities,
ii) calculations could also be performed at NNLO and iii) no assumptions about unfavoured FFs were imposed.
The quality of the fits were high
suggesting that perturbative QCD is consistent with the data, including the very low $x$ measurements.
However, the fitted kaon mass $m_K$, on which low $x$ cross sections depend strongly,
was found to be somewhat lower than the value obtained phenomenologically, i.e.\ without perturbative QCD, from
the fit to neutral kaon production data only.
The obtained values for $D_u^{K^\pm}-D_d^{K^\pm}$ at small $z$ are negative and considerably different from those obtained from
global analyses.
Fits performed using NNLO calculations gave lower theoretical errors on $D_u^{K^\pm}-D_d^{K^\pm}$, suggesting
stability of the perturbation series for the most of the available cross section measurements.

The current measurements of inclusive kaon production are not at the level of accuracy required to
obtain a competitive extraction of $\alpha_s(M_Z)$ from the considered cross section differences.
Also it is not enough to really constrain the non-singlet $D_{u-d}^{K^\pm}$.
As our fits show, the error on the FF is large (see Fig. 4).
But our method is a good one, which  allows for the first NNLO analysis of inclusive hadron production
and it will be useful for future studies.
Admittedly the experimental errors are large, still one would not get such good fits to cross section
(not cross section difference) measurements, even with similarly large errors,
due to small $x$-logarithms in singlet / gluon evolution and in gluon coefficient functions.

With more accurate data in the future and in greater number, it would be nice to see if we could continue
to describe the low $x$ data well.
In particular, at present the most accurate data is that for which $\sqrt{s}=91.2$ GeV,
which is the least sensitive
to the kaon non singlet due to the similarity between the $u$ and $d$ quark effective electroweak charges at this energy.
However, such an extraction may become possible once the accurate
measurements of kaon production at BaBar \cite{Anulli:2004nm} have been finalized,
because these data are at $\sqrt{s} =10.54$ GeV where the quark electroweak charges are very different.
These BaBar data could also significantly improve the constraints on $D_u^{K^\pm}-D_d^{K^\pm}$.

\appendix
\section{Harmonic polylogarithms in Mellin space}
\label{NNLOHS}

In this section we describe our procedure for obtaining the Mellin transform
of harmonic sums, defined as
\be
H_{m_1,m_2,\ldots,m_n}(x) = \int_0^x dx_1 f_{m_1}(x_1) \int_0^{x_1} dx_2 f_{m_2}(x_2) \ldots
\int_0^{x_{n-1}}dx_n f_{m_n}(x_n)
\label{defofHx}
\ee
where
\be
f_0(x)=\frac{1}{x},\qquad
f_1(x)=\frac{1}{1-x},\qquad
f_{-1}(x)=\frac{1}{1+x}
\ee
Their Mellin transforms can be expressed as a weighted sum of harmonic sums,
defined for integer values of the Mellin space variable $n$ as
\be
S_{k_1,k_2,k_3,\ldots}(n)
=\sum_{n_1=1}^n \frac{({\rm sgn}(k_1))^{n_1}}{n_1^{|k_1|}}\sum_{n_2=1}^{n_1}\frac{({\rm sgn}(k_2))^{n_2}}{n_2^{|k_2|}}
\sum_{n_3=1}^{n_2}\frac{({\rm sgn}(k_3))^{n_3}}{n_3^{|k_3|}}\ldots
\label{defofHS}
\ee
which can be continued to complex $n$ according to the procedure in \cite{Albino:2009ci}.

Such weighted sums can be found recursively \cite{Remiddi:1999ew} by determining, in Mellin space, the dependence of
a harmonic polylogarithm on the same one without the leftmost index.
By performing the integration for the Mellin transform of
$H_{p,\overrightarrow{m}}(x)=\int_0^x dy f_p(y) H_{\overrightarrow{m}}(y)$ by parts, which gives
$\widetilde{H}_{p,\overrightarrow{m}}(n)=(H_{p,\overrightarrow{m}}(1)-M\left[x f_p(x)H_{\overrightarrow{m}}(x)\right](n))/n$,
and then performing the replacements $xf_{\pm 1}(x)=\mp \left(1-f_{\pm 1}(x)\right)$, we find that our desired relations are
\be
\widetilde{H}_{0,\overrightarrow{m}}(n)=\frac{H_{0,\overrightarrow{m}}(1)-\widetilde{H}_{\overrightarrow{m}}(n)}{n}
\ee
\be
\widetilde{H}_{1,\overrightarrow{m}}(n)=\frac{\widetilde{H}_{\overrightarrow{m}}(n)
-M\left[\left[\frac{H_{\overrightarrow{m}}(x)}{1-x}\right]_+ \right](n)}{n}
\ee
\be
\widetilde{H}_{-1,\overrightarrow{m}}(n)=\frac{H_{-1,\overrightarrow{m}}(1)-\widetilde{H}_{\overrightarrow{m}}(n)
+M\left[\frac{H_{\overrightarrow{m}}(x)}{1+x}\right](n)}{n}
\ee
where $M\left[H_{\overrightarrow{m}}(x)/(1+x) \right](n)$ and the ``+'' distribution
$M\left[\left[H_{\overrightarrow{m}}(x)/(1-x)\right]_+ \right](n)
=M\left[H_{\overrightarrow{m}}(x)/(1-x)\right](n)-M\left[H_{\overrightarrow{m}}(x)/(1-x)\right](1)$
can be calculated from $\widetilde{H}_{\overrightarrow{m}}(n)$ by expanding $1/(1\pm x)$
as a series in $x$ before performing the Mellin transform.
The result \cite{Vermaseren:1998uu} is simply that, because
$\sum_{i=1}^n (\mp 1)^i S_{\overrightarrow{m}}(i)/i^p=S_{\mp p,\overrightarrow{m}}(n)$
which follows from eq.\ (\ref{defofHS}), where $p>0$ here and in what follows,
each term of the form
\be
\frac{S_{\overrightarrow{m}}(n)}{n^p}\qquad {\rm in}\qquad \widetilde{H}_{\overrightarrow{r}}(n)
\ee
becomes
\be
(\mp 1)^n\left[S_{\mp p,\overrightarrow{m}}(\infty)-S_{\mp p,\overrightarrow{m}}(n-1)\right]
=\frac{S_{\overrightarrow{m}}(n)}{n^p}
+(\mp 1)^n\left[S_{\mp p,\overrightarrow{m}}(\infty)-S_{\mp p,\overrightarrow{m}}(n)\right]
\qquad {\rm in}\qquad M\left[\frac{H_{\overrightarrow{r}}(x)}{1\pm x}\right](n)
\ee
Furthermore, each term of the form
\be
(-1)^n \frac{S_{\overrightarrow{m}}(n)}{n^p}\qquad {\rm in}\qquad \widetilde{H}_{\overrightarrow{r}}(n)
\ee
becomes
\be
\nonumber
(\mp 1)^n\left[S_{\pm p,\overrightarrow{m}}(\infty)-S_{\pm p,\overrightarrow{m}}(n-1)\right]
=(-1)^n \frac{S_{\overrightarrow{m}}(n)}{n^p}
+(\mp 1)^n\left[S_{\pm p,\overrightarrow{m}}(\infty)-S_{\pm p,\overrightarrow{m}}(n)\right]
\qquad {\rm in}\qquad M\left[\frac{H_{\overrightarrow{r}}(x)}{1\pm x}\right](n)\\
\ee
Although the $S_{1,\overrightarrow{m}}(\infty)$ are singular, they may be treated in a symbolic sense
because $\widetilde{H}_{\overrightarrow{m}}(n)$,
$M\left[H_{\overrightarrow{m}}(x)/(1+x) \right](n)$ and
$M\left[\left[H_{\overrightarrow{m}}(x)/(1-x)\right]_+ \right](n)$
are all finite when the real part of $n$ is suitably large.
To complete this recursive procedure, we require the Mellin transforms of the simplest harmonic polylogarithms,
which are given by
\be
\widetilde{H}_0(n)=-\frac{1}{n^2}, \qquad
\widetilde{H}_1(n)=\frac{S_1(n)}{n}, \qquad
\widetilde{H}_{-1}(n)=-(-1)^n \frac{S_{-1}(n)}{n}+\frac{\ln(2)}{n}(1-(-1)^n)
\ee

A Mathematica file for implementing this procedure and for producing FORTRAN programs
to calculate numerical values of harmonic polylogarithms anywhere in Mellin space
can be obtained from \verb+http://www.desy.de/~simon/HarmonicSums+.

\begin{acknowledgments}

The authors would like to thank E.\ Leader for a careful reading of the early draft of the
manuscript and for numerous discussions.
The work of E.\ C.\ was supported by HEPTools EU network  MRTN-CT-2006-035505 and
the Bulgarian National Science Foundation, Grant 288/2008.

\end{acknowledgments}

\end{document}

%% file: K0Stable.tex
TASSO \cite{Althoff:1984iz} & $d\sigma^{K_S^0}$ & untagged &  14.0 &    9 & 15    &     0.4 &    -0.6  \\
\hline
TASSO \cite{Braunschweig:1989wg} & $d\sigma^{K_S^0}$ & untagged &  14.8 &    9 &       &     0.3 &    -0.4  \\
\hline
TASSO \cite{Braunschweig:1989wg} & $d\sigma^{K_S^0}$ & untagged &  21.5 &    6 &       &     0.0 & \\
\hline
TASSO \cite{Althoff:1984iz} & $d\sigma^{K_S^0}$ & untagged &  22.0 &    6 &       &     0.1 &     0.1  \\
\hline
HRS \cite{Derrick:1985wd} & $d\sigma^{K_S^0}$ & untagged &    29 &   13 &       &     3.2 & \\
\hline
MARK II & $d\sigma^{K_S^0}$ & untagged &  29.0 &   21 & 12    &     0.8 &     0.3  \\
\hline
TPC \cite{Aihara:1984mk} & $d\sigma^{K_S^0}$ & untagged &    29 &    8 &       &     0.5 & \\
\hline
TASSO \cite{Brandelik:1981ta} & $d\sigma^{K_S^0}$ & untagged &  33.3 &    9 & 15    &     0.7 &     0.2  \\
\hline
TASSO \cite{Althoff:1984iz} & $d\sigma^{K_S^0}$ & untagged &  34.0 &   15 &       &     1.4 &    -0.1  \\
\hline
TASSO \cite{Braunschweig:1989wg} & $d\sigma^{K_S^0}$ & untagged &  34.5 &   15 &       &     1.3 & \\
\hline
CELLO \cite{Behrend:1989ae} & $d\sigma^{K_S^0}$ & untagged &    35 &   11 &       &     0.5 & \\
\hline
TASSO \cite{Braunschweig:1989wg} & $d\sigma^{K_S^0}$ & untagged &    35 &   15 &       &     1.3 & \\
\hline
TASSO \cite{Braunschweig:1989wg} & $d\sigma^{K_S^0}$ & untagged &  42.6 &   15 &       &     0.5 & \\
\hline
TOPAZ \cite{Itoh:1994kb} & $d\sigma^{K_S^0}$ & untagged &    58 &    7 &       &     0.1 & \\
\hline
ALEPH \cite{Barate:1996fi} & $d\sigma^{K_S^0}$ & untagged &  91.2 &   30 & 2     &     0.5 &    -2.3  \\
\hline
DELPHI \cite{Abreu:1994rg} & $d\sigma^{K_S^0}$ & untagged &  91.2 &   26 &       &     0.7 & \\
\hline
OPAL \cite{Abbiendi:2000cv} & $d\sigma^{K_S^0}$ & untagged &  91.2 &   20 & 6     &     1.0 &    -1.1  \\
\hline
OPAL \cite{Abbiendi:1999ry} & $d\sigma^{K_S^0}$ & c tagged &  91.2 &    5 &       &     0.6 & \\
\hline
OPAL \cite{Abbiendi:1999ry} & $d\sigma^{K_S^0}$ & b tagged &  91.2 &    5 &       &     1.7 & \\
\hline
SLD \cite{Abe:1998zs} & $d\sigma^{K_S^0}$ & untagged &  91.2 &   17 &       &     1.1 & \\
\hline
SLD \cite{Abe:1998zs} & $d\sigma^{K_S^0}$ & l tagged &  91.2 &   17 &       &     0.6 & \\
\hline
SLD \cite{Abe:1998zs} & $d\sigma^{K_S^0}$ & c tagged &  91.2 &   17 &       &     0.7 & \\
\hline
\hline
SLD \cite{Abe:1998zs} & $d\sigma^{K_S^0}$ & b tagged &  91.2 &   17 &       &     1.5 & \\
\hline
DELPHI \cite{Abreu:2000gw} & $d\sigma^{K_S^0}$ & untagged &   189 &   10 &       &     0.7 & \\
\hline
\hline
DELPHI \cite{Abreu:2000gw} & $d\sigma^{K_S^0}$ & untagged &   183 &    8 &       &     1.3 & \\
\hline
\hline
 &  &  &  &  331 &  &     1.1 & \\
\hline

%% file: Ktable.tex
TASSO \cite{Brandelik:1980iy} & $d\sigma^{K^\pm}$ & untagged &    12 &    3 & 20    &     1.0 &    -1.1 &     0.9 &    -1.0  \\
\hline
TASSO \cite{Althoff:1982dh} & $d\sigma^{K^\pm}$ & untagged &    14 &    9 & 8.5   &     1.0 &    -0.1 &     0.9 &    -0.2  \\
\hline
TASSO \cite{Althoff:1982dh} & $d\sigma^{K^\pm}$ & untagged &    22 &   10 & 6.3   &     0.3 &    -0.5 &     0.3 &    -0.6  \\
\hline
HRS \cite{Derrick:1985wd} & $d\sigma^{K^\pm}$ & untagged &    29 &    7 &       &     1.9 & &     2.3 & \\
\hline
MARKII \cite{Schellman:1984yz} & $d\sigma^{K^\pm}$ & untagged &    29 &    6 & 12    &     2.1 &    -1.4 &     2.6 &    -1.4  \\
\hline
TPC \cite{Aihara:1988fc} & $d\sigma^{K^\pm}$ & untagged &    29 &   29 &       &     1.2 & &     1.8 & \\
\hline
TASSO \cite{Brandelik:1980iy} & $d\sigma^{K^\pm}$ & untagged &    30 &    5 & 20    &     0.9 &    -1.4 &     0.9 &    -1.4  \\
\hline
TASSO \cite{Braunschweig:1988hv} & $d\sigma^{K^\pm}$ & untagged &    34 &   11 & 6     &     1.5 &    -1.0 &     1.6 &    -0.9  \\
\hline
TASSO \cite{Braunschweig:1988hv} & $d\sigma^{K^\pm}$ & untagged &    44 &    4 & 6     &     0.1 & &     0.1 & \\
\hline
TOPAZ \cite{Itoh:1994kb} & $d\sigma^{K^\pm}$ & untagged &    58 &   12 &       &     0.7 & &     0.7 & \\
\hline
ALEPH \cite{Buskulic:1994ft,Barate:1996fi} & $d\sigma^{K^\pm}$ & untagged &  91.2 &   29 & 3     &     1.3 &    -0.6 &     1.2 &    -0.7  \\
\hline
DELPHI \cite{Abreu:1998vq} & $d\sigma^{K^\pm}$ & untagged &  91.2 &   23 &       &     0.2 & &     0.2 & \\
\hline
DELPHI \cite{Abreu:1998vq} & $d\sigma^{K^\pm}$ & l tagged &  91.2 &   23 &       &     0.8 & &     0.8 & \\
\hline
DELPHI \cite{Abreu:1998vq} & $d\sigma^{K^\pm}$ & b tagged &  91.2 &   23 &       &     0.5 & &     0.5 & \\
\hline
OPAL \cite{Akers:1994ez} & $d\sigma^{K^\pm}$ & untagged &  91.2 &   33 &       &     2.3 & &     2.5 & \\
\hline
OPAL \cite{Abbiendi:1999ry} & $d\sigma^{K^\pm}$ & c tagged &  91.2 &    5 &       &     4.6 & &     4.8 & \\
\hline
OPAL \cite{Abbiendi:1999ry} & $d\sigma^{K^\pm}$ & b tagged &  91.2 &    5 &       &     4.5 & &     4.4 & \\
\hline
SLD \cite{Abe:2003iy} & $d\sigma^{K^\pm}$ & untagged &  91.2 &   36 &       &     1.9 & &     1.5 & \\
\hline
SLD \cite{Abe:2003iy} & $d\sigma^{K^\pm}$ & l tagged &  91.2 &   36 &       &     4.7 & &     4.0 & \\
\hline
SLD \cite{Abe:2003iy} & $d\sigma^{K^\pm}$ & c tagged &  91.2 &   36 &       &     2.7 & &     2.4 & \\
\hline
SLD \cite{Abe:2003iy} & $d\sigma^{K^\pm}$ & b tagged &  91.2 &   36 &       &     4.7 & &     4.6 & \\
\hline
DELPHI \cite{Abreu:2000gw} & $d\sigma^{K^\pm}$ & untagged &   189 &    8 &       &     5.2 & &     5.3 & \\
\hline
OPAL \cite{Abbiendi:1999ry} & $d\sigma^{K^\pm}-2d\sigma^{K_S^0}$ & u tagged &  91.2 &    5 &       &     1.2 & &     1.4 & \\
\hline
OPAL \cite{Abbiendi:1999ry} & $d\sigma^{K^\pm}-2d\sigma^{K_S^0}$ & d tagged &  91.2 &    5 &       &     1.0 & &     1.5 & \\
\hline
TASSO \cite{Althoff:1984iz} & $d\sigma^{K_S^0}$ & untagged &    14 &    9 & 15    &     0.4 &    -0.2 &     0.4 &     0.0  \\
\hline
TASSO \cite{Braunschweig:1989wg} & $d\sigma^{K_S^0}$ & untagged &  14.8 &    9 &       &     0.6 & &     0.6 & \\
\hline
TASSO \cite{Braunschweig:1989wg} & $d\sigma^{K_S^0}$ & untagged &  21.5 &    6 &       &     0.1 & &     0.1 & \\
\hline
TASSO \cite{Althoff:1984iz} & $d\sigma^{K_S^0}$ & untagged &    22 &    6 &       &     0.2 &     0.2 &     0.3 &     0.3  \\
\hline
HRS \cite{Derrick:1985wd} & $d\sigma^{K_S^0}$ & untagged &    29 &   13 &       &     2.9 & &     3.4 & \\
\hline
MARK II & $d\sigma^{K_S^0}$ & untagged &    29 &   21 & 12    &     1.2 &     1.2 &     1.3 &     1.4  \\
\hline
TPC \cite{Aihara:1984mk} & $d\sigma^{K_S^0}$ & untagged &    29 &    8 &       &     1.8 & &     2.3 & \\
\hline
TASSO \cite{Brandelik:1981ta} & $d\sigma^{K_S^0}$ & untagged &  33.3 &    9 & 15    &     0.6 &     0.3 &     0.7 &     0.4  \\
\hline
TASSO \cite{Althoff:1984iz} & $d\sigma^{K_S^0}$ & untagged &    34 &   15 &       &     1.3 &     0.0 &     1.3 &     0.0  \\
\hline
TASSO \cite{Braunschweig:1989wg} & $d\sigma^{K_S^0}$ & untagged &  34.5 &   15 &       &     1.3 & &     1.2 & \\
\hline
CELLO \cite{Behrend:1989ae} & $d\sigma^{K_S^0}$ & untagged &    35 &   11 &       &     0.6 & &     0.6 & \\
\hline
TASSO \cite{Braunschweig:1989wg} & $d\sigma^{K_S^0}$ & untagged &    35 &   15 &       &     1.9 & &     1.9 & \\
\hline
TASSO \cite{Braunschweig:1989wg} & $d\sigma^{K_S^0}$ & untagged &  42.6 &   15 &       &     0.6 & &     0.6 & \\
\hline
TOPAZ \cite{Itoh:1994kb} & $d\sigma^{K_S^0}$ & untagged &    58 &    7 &       &     1.1 & &     1.0 & \\
\hline
ALEPH \cite{Barate:1996fi} & $d\sigma^{K_S^0}$ & untagged &  91.2 &   30 & 2     &     1.5 &     1.9 &     1.4 &     1.7  \\
\hline
DELPHI \cite{Abreu:1994rg} & $d\sigma^{K_S^0}$ & untagged &  91.2 &   26 &       &     3.0 & &     2.8 & \\
\hline
OPAL \cite{Abbiendi:2000cv} & $d\sigma^{K_S^0}$ & untagged &  91.2 &   20 & 6     &     2.3 &     0.4 &     2.2 &     0.3  \\
\hline
OPAL \cite{Abbiendi:1999ry} & $d\sigma^{K_S^0}$ & c tagged &  91.2 &    5 &       &     1.2 & &     1.3 & \\
\hline
OPAL \cite{Abbiendi:1999ry} & $d\sigma^{K_S^0}$ & b tagged &  91.2 &    5 &       &    12.9 & &    12.8 & \\
\hline
SLD \cite{Abe:1998zs} & $d\sigma^{K_S^0}$ & untagged &  91.2 &   17 &       &     3.2 & &     3.0 & \\
\hline
SLD \cite{Abe:1998zs} & $d\sigma^{K_S^0}$ & l tagged &  91.2 &   17 &       &     0.8 & &     0.7 & \\
\hline
SLD \cite{Abe:1998zs} & $d\sigma^{K_S^0}$ & c tagged &  91.2 &   17 &       &     1.2 & &     1.2 & \\
\hline
SLD \cite{Abe:1998zs} & $d\sigma^{K_S^0}$ & b tagged &  91.2 &   17 &       &     5.1 & &     5.2 & \\
\hline
DELPHI \cite{Abreu:2000gw} & $d\sigma^{K_S^0}$ & untagged &   183 &    8 &       &     1.9 & &     1.9 & \\
\hline
DELPHI \cite{Abreu:2000gw} & $d\sigma^{K_S^0}$ & untagged &   189 &   10 &       &     2.7 & &     2.7 & \\
\hline
\hline
 &  &  &  &  730 &  &     2.3 & &     2.2 & \\
\hline